\documentclass[iop]{emulateapj}
\pdfoutput=1
\usepackage{amsmath}
\usepackage{hyperref}
\usepackage{graphicx}
\usepackage{grffile}
\newcommand*\mean{\overline}
\newcommand{\cmb}{{\rm CMB}}
\newcommand{\noise}{{\rm noise}}
\newcommand{\gal}{{\rm gal}}

\newcommand{\pix}{{\rm pix}}
\newcommand{\tfoo}{\theta_{500}}

\begin{document}

\title{Optimizing measurements of cluster velocities and temperatures
for CCAT-prime and future surveys}

\author{Avirukt Mittal, Francesco de Bernardis and Michael D. Niemack}
\affil{Department of Physics, Cornell University,
    Ithaca, NY 14853}

\begin{abstract}
Galaxy cluster velocity correlations and mass distributions are sensitive probes of cosmology and the growth of structure. Upcoming microwave surveys will enable extraction of velocities and temperatures from many individual clusters for the first time. We forecast constraints on peculiar velocities, electron temperatures, and optical depths of galaxy clusters obtainable with upcoming multi-frequency measurements of the kinematic, thermal, and relativistic Sunyaev-Zeldovich effects. The forecasted constraints are compared for different measurement configurations with frequency bands between 90 GHz and 1 THz, and for different survey strategies for the 6-meter CCAT-prime telescope. We study methods for improving cluster constraints by removing emission from dusty star forming galaxies, and by using X-ray temperature priors from eROSITA. Cluster constraints are forecast for several model cluster masses. A sensitivity optimization for seven frequency bands is presented for a CCAT-prime first light instrument and a next generation instrument that takes advantage of the large optical throughput of CCAT-prime. We find that CCAT-prime observations are expected to enable measurement and separation of the SZ effects to characterize the velocity, temperature, and optical depth of individual massive clusters ($\sim10^{15}\, M_\odot$). Submillimeter measurements are shown to play an important role in separating these components from dusty galaxy contamination. Using a modular instrument configuration with similar optical throughput for each detector array, we develop a rule of thumb for the number of detector arrays desired at each frequency to optimize extraction of these signals. Our results are relevant for a future ``Stage IV" cosmic microwave background survey, which could enable galaxy cluster measurements over a larger range of masses and redshifts than will be accessible by other experiments.
\end{abstract}

\section{Introduction}

Galaxy cluster measurements have played an important role in establishing the dark energy and dark matter dominated cosmological model \citep[e.g.,][]{allen/etal:2011,planck:2015sz}. 
Future measurements of the peculiar velocities of galaxy clusters will probe physics on large scales and have the potential to place strong constraints on cosmological parameters, complementary to those achievable with measurements of the density field \citep{1994ApJ...436...23B,1994ApJ...437L..51C,1994MNRAS.268L..23C,1996ApJ...462L..49B,1996MNRAS.282..384M,2008PhRvD..77h3004B,2015ApJ...808...47M,2015PhRvD..92f3501M}.

However, measuring peculiar velocities is a difficult task. The Sunyaev-Zeldovich (SZ) effects \citep{1970Ap&SS...7....3S,1972CoASP...4..173S} offer a promising approach for measuring peculiar velocities. Photons from the Cosmic Microwave Background (CMB) interact with hot electron gas in the intracluster medium (ICM). Through inverse Compton scattering, the electrons boost the photon energy, distorting the CMB blackbody spectrum when observed in the direction of a galaxy cluster. The SZ effects consist of a thermal (tSZ) component related to the thermal energy of the scattering electrons, a relativistic (rSZ) component also related to the electron temperature, and a kinematic (kSZ) component related to the bulk motion of the electrons \citep[e.g.,][]{birkinshaw:1999}. The kSZ effect is generally more than ten times smaller than the tSZ effect for massive clusters ($M>10^{14}M_{\odot}$), and its amplitude is proportional to the peculiar velocity of the cluster along the line of sight of the observer. 

While the tSZ effect has been measured for well over a thousand clusters \citep[e.g.,][]{2013JCAP...07..008H,2015ApJS..216...27B,planck:2015sz}, measurements of the kSZ for individual clusters remain difficult to achieve. \cite{2012ApJ...761...47M} measured the peculiar velocity of one of the merging clusters within MACS J0717.5+3745 to be $v = 3450\pm900 \rm \:km\,s^{-1}$. \cite{sayers/etal:2013} measure the kSZ effect of the same cluster to higher significance, and \cite{2017A&A...598A.115A} find a statistically significant kSZ dipole in the merging system. Similarly, while preliminary evidence for the rSZ effect exists from one individual cluster \citep{zemcov/etal:2012} and one analysis of stacked clusters \citep{2016A&A...596A..61H}, measurements have not been sufficiently sensitive to extract the rSZ effect from multiple individual clusters thus far.

In the last few years several groups have pursued measuring the kSZ effect with a statistical approach over a large sample of clusters using several estimators. The first detection of this kind was achieved by \cite{2012PhRvL.109d1101H} with a pairwise kSZ estimator. Detections using similar estimators have subsequently been reported by \cite{Ade:2015lza,2016MNRAS.461.3172S,2016arXiv160702139D}. An alternative estimator based on correlating a velocity template with CMB temperature maps was used in
\cite{Schaan:2015uaa} and \cite{Ade:2015lza} for which the velocity template was constructed from measurements of the large-scale density field assuming the continuity equation.
\cite{Hill:2016dta} used squared CMB anisotropy maps 
cross-correlated with galaxy measurements 
as another means of extracting statistical evidence for the kSZ effect. \citet{planck_ksz:2017} used tSZ cleaned maps to measure the velocity dispersion of a large sample of X-ray detected clusters.

These statistical measurements are promising and contain significant cosmological information, though the statistical power is not yet sufficiently high to provide competitive cosmological constraints. As the measurements improve with upcoming surveys, cosmological constraints should be possible, provided that the tSZ effect can be removed effectively and the cluster optical depth (the electron gas density integrated along the line of sight) can be estimated independently or marginalized over \citep{2015ApJ...808...47M,2015PhRvD..92f3501M,Ferraro:2016ymw}. 

In this paper we use a Fisher matrix approach to explore the ability of upcoming multi-frequency surveys to measure the kSZ effect and line-of-sight peculiar velocity for individual clusters, by separating the tSZ and rSZ components. The three SZ components depend on the electron temperature ($T_e$), optical depth ($\tau$), and the peculiar velocity ($v$). 
Measuring $v$, $T_e$, and $\tau$ directly for a large sample of galaxy clusters will enable use of new kSZ statistics for constraining cosmology \citep[e.g,][]{2008PhRvD..77h3004B} and will reduce systematic effects, such as residual tSZ signal and unknown optical depth, which may limit the potentially powerful statistics described above. However, approaches based on cross-correlations with the galaxy field are less affected by other sources of emission, such as emission from dusty star-forming galaxies (DSFGs). In this analysis DSFGs and the CMB represent significant sources of noise for direct measurements of the kSZ, rSZ, and tSZ effects. 
With the recent progress in SZ measurements it is timely to study and optimize the potential of upcoming multi-frequency surveys to separate the SZ signals from these other astrophysical sources.

We update and extend the forecasts presented in \cite{knox} by using a universal pressure profile, including new estimates for dusty galaxy contamination, assigning foreground measurement bands to isolate the dusty galaxy contamination, and applying the forecasts to a realistic distribution of cluster parameters observable with upcoming surveys. We focus on 
the 6 meter CCAT-prime (CCAT-p) telescope,\footnote{\url{http://www.ccatobservatory.org/}} which is expected to begin first light observations in 2021. We also consider first light and next generation CCAT-p measurements in combination with those from other experiments, such as eROSITA \citep{2010SPIE.7732E..0UP,2012arXiv1209.3114M} and Advanced ACTPol \citep{Henderson:2015nzj}. The fiducial cosmological model assumed throughout this paper is based on \cite{2015arXiv150201589P}.

A summary of the SZ effects is presented in \S \ref{SZ}. \S \ref{fisher} describes the Fisher matrix method used for the analysis,and the figure of merit used to compare forecast results is described in \S \ref{fom}. \S\ref{noise} presents models of SZ contamination sources.  \S\ref{foreground} describes the DSFG foreground removal approach. \S \ref{experiments} describes the experimental parameters used for the forecasts. Forecast results for different CCAT-p configurations are presented in \S \ref{results}, followed by conclusions in \S\ref{conclusion}. Appendix  \ref{sec:comparison} compares our forecasts with real data from existing surveys analyzed by \cite{lindner} and discusses possible systematics. The map pixelization used for the forecasts is discussed in Appendix \ref{sec:pixel}.

\section{The Sunyaev-Zeldovich Effects}\label{SZ}

The amplitude of the tSZ effect at the cluster center, known as the comptonization parameter, is
\begin{equation}
y\equiv\int \sigma_T n_e \Theta \:dl,
\end{equation}
where $\sigma_T$ is the Thomson scattering cross-section, $n_e$ is the electron number density, $l$ represents the line of sight through the center of the cluster, and
\begin{equation*}
\Theta \equiv \frac{kT_e}{m_ec^2}
\end{equation*}
is the dimensionless electron gas temperature. 
The comptonization parameter can be expressed in terms of the optical depth, $\tau$, as
\begin{equation}
y = \tau \mean{\Theta},
\end{equation}
where
\begin{equation}
\tau\equiv\int \sigma_T n_e \:dl
\end{equation}
and the bar represents an average, weighted by optical depth (i.e. by density), along the line of sight through the cluster center.

Throughout this paper, we work in units of CMB brightness temperature. To convert between intensity and CMB temperature, we linearize Planck's law using the first-order Taylor expansion in temperature:
\begin{equation}\label{Jy_conv}
\begin{split}
\frac{\partial T^\cmb}{\partial B^\cmb_\nu} &= \frac{c^2}{2k}\left(\frac{x}{\nu}\right)^2\frac{(e^x - 1)^2}{x^4\,e^x}\\
&= \frac{119\:\mathrm{\mu K}}{1\:\mathrm{mJy}/(1'\times 1')}\frac{(e^x - 1)^2}{x^4\,e^x}\\
\end{split}
\end{equation}

where the spectral intensity $B_\nu$ is given by Planck's law and $x$ is the dimensionless frequency, $x \equiv h \nu/k T_\mathrm{CMB}$. For convenience we also define $\tilde{x} \equiv x \, \mathrm{coth}(x/2).$

The frequency dependence of the tSZ effect can be written as
\begin{equation}
f_1(\nu) = x \, \mathrm{coth}(x/2) - 4 \equiv \tilde{x} - 4
\end{equation}
\citep[e.g.][]{itoh}. Since the electron gas in massive clusters is typically at high temperatures ($\sim \rm1\: keV\approx10^7\:K$), the relativistic correction to the tSZ effect, which is higher-order in $\Theta$, is non-negligible. The first-order rSZ correction has an amplitude equal to
\begin{equation}
\int \sigma_T n_e \Theta^2 \:dl=\int_l \Theta^2\: d\tau = \tau \mean{\Theta^2}
\end{equation}
and a frequency dependence modeled as
\begin{equation}
\begin{split}
f_2(\nu) =& -10 + 23.5\,\tilde{x} - 8.4\,\tilde{x}^2 +0.7\,\tilde{x}^3 \\
&+ (-4.2+1.4\,\tilde{x})\left(\frac{x}{\mathrm{sinh}(x/2)}\right)^2.
\end{split}
\end{equation}
While higher order rSZ corrections must be accounted for when measuring parameters in the most massive clusters, this approximation is expected to be valid for clusters up to 10 keV with corrections smaller than five percent \citep{itoh}.

\begin{figure}
\includegraphics[width=\columnwidth]{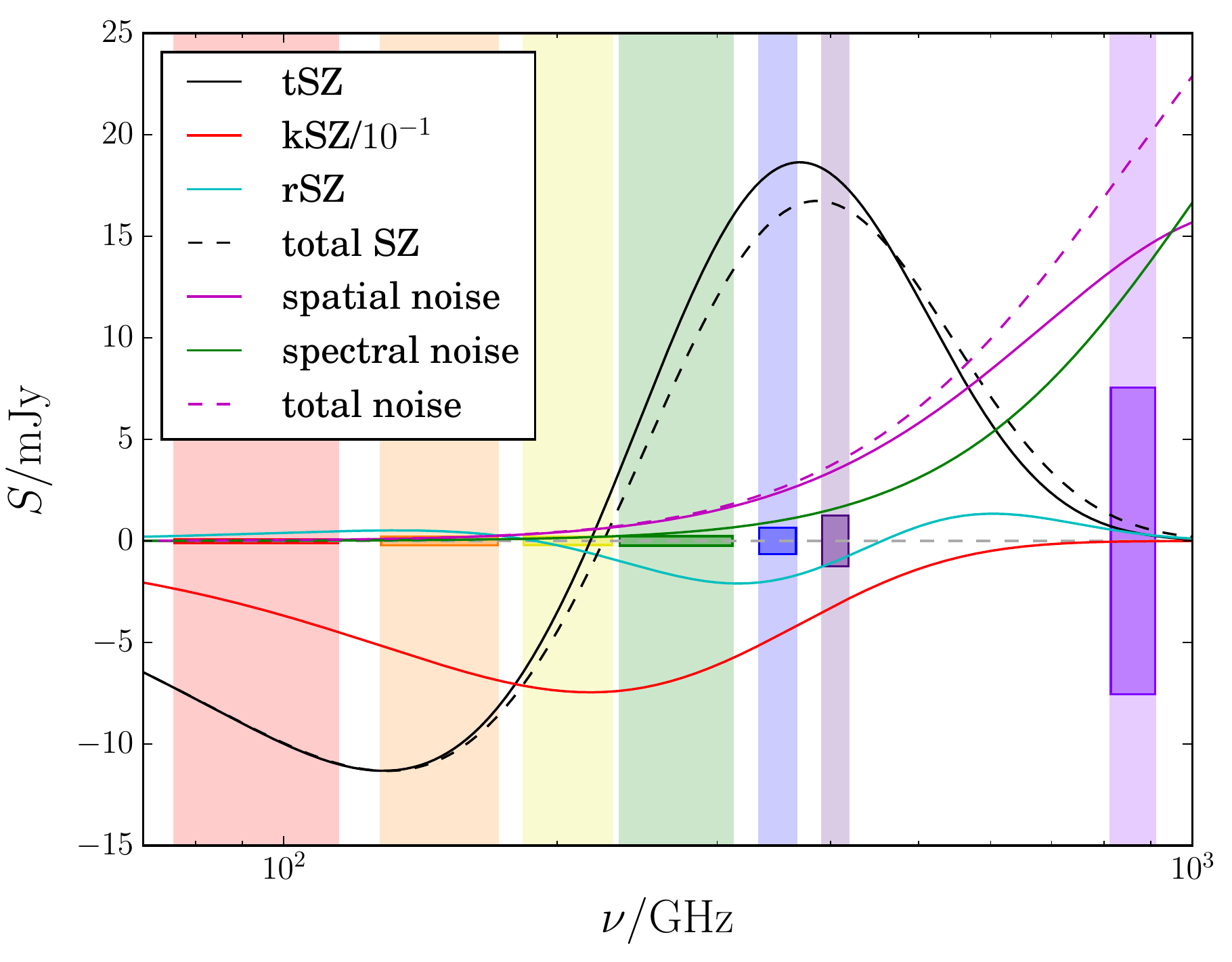}
\includegraphics[width=\columnwidth]{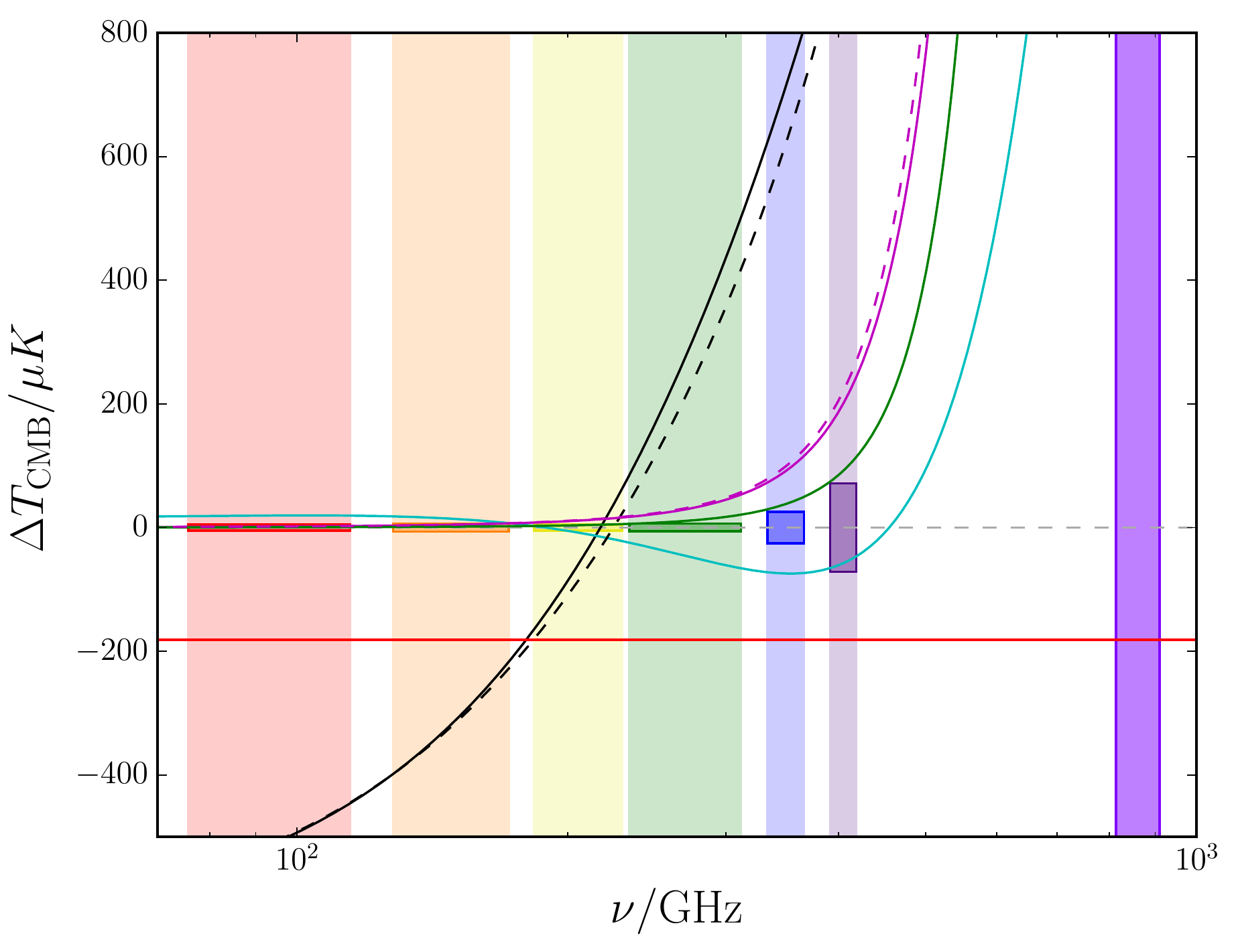}
\caption{Frequency spectra of the various SZ and noise components for the fiducial cluster described by Table \ref{fiducial_cluster}. The kSZ signal has been amplified by a factor of 10 for clarity.
The vertical color bars show the seven frequency bands being studied for CCAT-p, and the darkened regions indicate map noise estimates from Table~\ref{expt_param}.
``Noise" refers to the expected standard deviation in galactic noise, since no other noise has a  frequency dependent temperature. The noise spatial term corresponds to $f_4$, while the spectral term corresponds to $f_5$ (eq. \ref{t_gal_eq}). All beam and pixel solid angles are assumed to be one square arcminute 
for these plots. 
\textit{Top:} The frequency spectra in intensity, $S$, relative to the CMB blackbody. \textit{Bottom:} The same spectra in CMB-equivalent brightness temperature. The lines have the same meaning as the top panel
\label{fig:frequency_dist}}
\end{figure}

The kSZ effect is independent of frequency in CMB temperature units (and thus spectrally indistinguishable from the CMB anisotropies), since it results from the Doppler shift due to the motion of the scattering frame relative to the CMB rest frame; thus $f_3(\nu) = 1$. Its amplitude is given by
\begin{equation}
\int \sigma_Tn_e \frac{v}{c}\:dl = \int_l \frac{v}{c}\:d\tau = \tau \frac{\mean{v}}{c},
\end{equation}
where $v/c$ is the dimensionless velocity of the galaxy cluster relative to the CMB rest frame, i.e. its peculiar velocity. This effect is usually dominated by the tSZ effect (ref. Figure~\ref{fig:frequency_dist}). 

We assume that the clusters are isothermal and can be treated in the limit of an ideal electron gas, where the number density of electrons is traced by its pressure. This leads to a straightforward application of the universal pressure profile \citep[UPP,][]{arnaud} to all three SZ components. The intra-cluster velocity dispersion is ignored. 
The UPP is given by
\begin{equation}
\frac{P(xR_{500})}{P_0} = (c_{500}x)^{-\gamma}(1+(c_{500}x)^\alpha)^{(\gamma-\beta)/\alpha},
\end{equation}
where $R_{500}$ is the maximum cluster radius inside which the average density is 500 times the critical density of the universe. The quantities $\alpha,\beta,\gamma,c_{500}$ are the best-fit parameters given in \cite{arnaud} based on the analysis of 33 clusters from the Representative XMM-Newton Cluster Structure Survey (REXCESS): $\alpha = 1.051$, $\beta = 5.4905$, $\gamma = 0.3081$, $c_{500} = 1.177$.

The spatial variation of the cluster signal is given by
\begin{equation}\label{cluster_profile}
\begin{split}
h(\theta) &= A\int P(r)\:dl\\ &=  A\int_{-\infty}^\infty P\left(\sqrt{\left( \theta\,\frac{ R_{500}}{\tfoo}\right)^2 + l^2}\right ) dl,
\end{split}
\end{equation}
where $l$ represents the line of sight an angle $\theta$ away from the cluster center, $r$ is the radial distance to the center, $\tfoo$ is the angle corresponding to $R_{500}$ (which can be obtained with the cluster's redshift), and $A$ is a normalisation constant chosen such that $h(0)=1$ at the cluster center. An example cluster profile is shown in Figure \ref{int_signal_plot}, along with the resulting integrated comptonization parameter (equation \ref{int_y_eq}).

The complete SZ effect signal at frequency $\nu$ and pixel position $\pmb{\theta}$ in the sky relative to the cluster center can therefore be written using the frequency dependencies above with the amplitudes, $P_i$, of the tSZ, rSZ, and kSZ effects
\begin{equation}\label{p1p2p3}
P_1\equiv y =\tau \Theta,\hspace{11pt}P_2 \equiv \tau \Theta^2,\hspace{11pt}P_3 \equiv \tau \frac{v}{c},
\end{equation}
respectively, as
\begin{equation}\label{sz_signal}
\frac{\Delta T_{\rm SZ}(\nu, \pmb{\theta})}{T_\cmb} = (P_1 f_1(\nu) + P_2 f_2(\nu) + P_3 f_3(\nu))\, \tilde{h}(|\pmb{\theta}|),
\end{equation}
where $\tilde{h}(\theta)$ is the cluster profile convolved with the beam. Note that we have used the assumption of isothermality so that $\mean{\Theta^2} = \mean{\Theta}^2$ and defined $\Theta \equiv \mean{\Theta},\;v\equiv \mean{v}$ for brevity. The SZ signals for a fiducial cluster (whose parameters are given in Table~\ref{fiducial_cluster}) are presented in Figure~\ref{fig:frequency_dist}, along with the bands of CCAT-p and the corresponding baseline noise levels (ref. Table~\ref{expt_param}) for comparison. 

To assess the impact of our assumption of isothermality on the results, we consider a simple power law temperature profile, $T_e \propto r^{-0.24}$, from \cite{2008A&A...486..359L} \citep[which seems to agree with][at least for the range $0.2 R_{180} \leq r \leq 0.5 R_{200}$]{2007A&A...461...71P}. For example, Figure~\ref{int_signal_plot} depicts the spatial profiles of the three SZ components using the power law temperature profile. The plot also shows that the total kSZ signal of a non-isothermal cluster (within a reasonable aperture) is 50--70\% greater than that of an isothermal one, and the total rSZ signal is smaller by 30--40\%. Thus one would expect our isothermal forecasts to be pessimistic for kSZ and optimistic for rSZ uncertainties. Indeed, we find that using the power law temperature profile, the kSZ uncertainties are 30-40\% smaller, and the rSZ uncertainties bigger by a similar fraction, than those assuming isothermality. The effects of non-isothermality on the results are described briefly in \S\ref{ind_res}.

It is worth noting that the assumption of a specific temperature profile affects the relationship between $\mean{\Theta^2}$ and $\mean{\Theta}^2$ and thus determines the amplitude of the rSZ effect $P_2$. In addition, temperature profiles seem to display significantly more variance than pressure profiles do \citep{arnaud}, so the assumption of a specific temperature profile may introduce more bias and error in real measurements than that of a specific pressure profile.

\begin{deluxetable}{ccccccc}
\tablecaption{Fiducial cluster parameters used in Figure~\ref{fig:frequency_dist} and Appendix \ref{sec:pixel}. This roughly corresponds to a cluster with a mass of nearly $10^{15} M_\odot$ and redshift $z \approx 0.5$ (ref. \S\ref{results}). \label{fiducial_cluster}}
\tablehead{\colhead{$y$} & \colhead{$\Theta$} & \colhead{$T_e$} & \colhead{$\tau$} & \colhead{$v$} & \colhead{$\tfoo$}}
\startdata
$1.2\times10^{-4}$ & 0.012 & $6\:\mathrm{keV}$ & 0.01 & $-200\:\mathrm{km\,s^{-1}}$ & $3'$
\enddata
\end{deluxetable}

\section{Fisher matrix method}\label{fisher}
To forecast the constraints on cluster parameters we adopt a formalism and method similar to the one described in \cite{knox}.
The quantities $P_1, P_2, P_3$ defined in equation \ref{p1p2p3} are a natural choice for the free parameters of the problem, 
and thus the Fisher matrix entries corresponding to $P_1, P_2, P_3$, are independent of cluster physical parameters (except $\tfoo$). To transform the Fisher matrix from the parameters $P_1,P_2,P_3$ to the parameters $\Theta, \tau, v/c$ we use the transformation matrix
\begin{equation}\label{trans_matrix}
R_{l i} = \frac{\partial P_i}{\partial P_l} =\\
\begin{bmatrix}
\tau & 2\tau\Theta & 0\\
\Theta & \Theta^2 & v/c\\
0 & 0 & \tau
\end{bmatrix},
\end{equation}
where the index $i$ runs over the old parameter space and $\ell$ is the index for the new parameters.  The transformation of the Fisher matrix is then given by
\begin{equation}\label{transformation}
F'_{l m} = R_{l i} F_{ij} R_{mj} = (RFR^T)_{l m}.
\end{equation}
Therefore the covariance matrix $C=F^{-1}$ transforms as $C'=R'CR'^T$ where
\begin{equation}\label{cov_trans_matrix}
R' = (R^{-1})^T =\frac{1}{y}\\
\begin{bmatrix}
-\Theta & 1 & 0\\
2\tau & -\tau\Theta^{-1} & 0\\
-2v/c & \Theta^{-1}v/c & \Theta
\end{bmatrix}.
\end{equation}

A typical SZ survey will provide maps of the CMB sky at several frequencies. Following \cite{knox}, we treat each map pixel at each frequency as an observable and characterize the probability distribution of the observed temperature deviations $\Delta T_\alpha$ with a noise covariance matrix $C^\noise_{\alpha\beta}$, where the indices $\alpha$ and $\beta$ each run over the combinations of pixels and frequency channels of the experiment. We assume an explicit Cartesian grid for the observations, and the dependence of the results on the parameters of this grid are discussed in Appendix \ref{sec:pixel}.
Under these assumptions, the Fisher matrix element corresponding to parameters $P_i$ and $P_j$ is
\begin{equation}\label{fisher_maps}
F_{ij} = \frac{1}{T_\cmb^2}\frac{\partial \Delta T_\alpha}{\partial P_i}(C^\noise)^{-1}_{\alpha\beta}\frac{\partial \Delta T_\beta}{\partial P_j}.
\end{equation}

We recall that, in the Fisher matrix approach, the covariance between observed values of parameters $P_i$ and $P_j$ is $C_{ij} = F_{ij}^{-1}$.
Specifically, an estimate for the standard deviation of parameter $P_i$ is $\sigma(P_i)= \sqrt{C_{ii}}$ and represents a lower limit for the 1-$\sigma$ uncertainty in $P_i$ marginalized over the other free parameters of the model. Note that equation \ref{cov_trans_matrix} implies that $\sigma(\Theta)$ and $\sigma(\tau)$ are independent of $v$.

One advantage of the Fisher matrix method is that priors on the parameters can be included very easily by a simple summation of the corresponding parameter entries. For example, if $P_i$ has already been measured with a 1-$\sigma$ uncertainty $\sigma(P_i)$, then we can capture that information by adding it to the Fisher matrix as follows:
\begin{equation}
F^{\rm posterior}_{ii} = F_{ii} + F_{ii}^{\rm prior} = F_{ii} + \sigma(P_i)^{-2}.
\end{equation}
This is used to include temperature priors obtained from X-ray measurements of galaxy clusters in some forecasts below.

\section{Figure of merit} \label{fom} 

A useful way to characterize the simultaneous bounds on a combination of parameters from these surveys is by specifying a figure of merit (FoM). The FoM between a set of parameters is defined as the inverse area of the 1-$\sigma$ confidence ellipse in these parameters. Thus the FoM for a set of parameters is related to the determinant of the Fisher matrix marginalized over the other parameters, which is the inverse of the determinant of the corresponding submatrix of the covariance matrix \citep{2001PhRvD..64l3527H}. For parameters $P_1,\ldots,P_n$, the FoM is given by
\begin{equation}
{\rm FoM}(P_1,\ldots,P_n) = \left (\det\, [C_{1,\ldots,n}] \right )^{-\frac{1}{2}},
\end{equation}
and for all parameters this reduces to FoM $= \sqrt{\det\, F}$. Higher FoMs correspond to better measurements and smaller uncertainties on the parameters of interest. 

The choice of parameters for a FoM is not unique and can be subjective, but it is generally a useful approach for comparing multiple bounds simultaneously.
For this paper, we use the FoM in the parameters $\Theta,\tau,v/c$, given by
\begin{equation}\label{fom_trans}
{\rm FoM}(\Theta,\tau,v/c) = |\!\det R |\, (\det F)^{\frac{1}{2}} = y^2\, (\det F)^{\frac{1}{2}},
\end{equation}
where $R$ is the transformation matrix given in equation \ref{trans_matrix}, $F$ is the Fisher matrix in $P_1,P_2,P_3$ as defined in equation \ref{p1p2p3}, and $y$ is the comptonization parameter. It also follows from this that the FoM in any two equivalent parameterizations (such as $P_1, P_2, P_3$ versus $T_e, \tau, v$) are linearly related to each other, so when comparing FoMs for the same cluster either parameterization can be used.

\section{Sources of SZ contamination}\label{noise}
We consider multiple sources of SZ signal contamination, including CMB temperature anisotropies, instrument noise, and emission from dusty galaxies, and we discuss other potential sources of contamination in \S\ref{sec:othersources}. We quantify the three primary contaminants as contributing to a total noise term in our forecasts,
\begin{equation}
\Delta T_\noise = \Delta T_\cmb + \Delta T_{\rm ins} + \Delta T_\gal.
\end{equation}
Since the components are independent, the total noise covariance between two observations separates into the sum of the covariances of the components. Thus the covariance between the pixel at frequency $\nu_a$ and position $\pmb{\theta}_i$, represented by the index $\alpha$, and the pixel at frequency $\nu_b$ and position $\pmb{\theta}_j$, represented by the index $\beta$, is
\begin{equation}
\begin{split}
C^\noise_{\alpha\beta} &\equiv \frac{1}{T_\cmb^2}\langle \Delta T_\noise(\nu_a, \pmb{\theta}_i)\, \Delta T_\noise(\nu_b, \pmb{\theta}_j) \rangle \\
&= C^\cmb_{\alpha\beta} + C^{\rm ins}_{\alpha\beta} + C^{\rm gal}_{\alpha\beta}
\end{split}
\end{equation}
Each of these components is described below. 

\subsection{Cosmic Microwave Background anisotropies}

The relevant CMB anisotropies are fluctuations in the blackbody temperature of the CMB, which like the kSZ signal are independent of frequency in these units. 
We express the CMB covariance in terms of the well-known CMB power spectrum $C_\ell \equiv 1/(2\ell+1)\, \sum_m \langle|a_{\ell m}|^2\rangle$ and the full widths at half maximum (FWHMs) of the beams $\xi_a$ and $\xi_b$ at frequencies $\nu_a$ and $\nu_b$. The covariance of the CMB anisotropies between observations $\alpha$ and $\beta$ is given by
\begin{equation}
C_{\alpha\beta}^\cmb= \sum_\ell \frac{2\ell+1}{4\pi}C_\ell P_\ell\left({\rm cos}\, |\pmb{\theta}_i - \pmb{\theta}_j|\right){\rm exp} \left (-\frac{\ell^2 (\xi_a^2+\xi_b^2)}{16\, {\rm ln}\, 2}\right).
\end{equation}
All beams are assumed to be Gaussian throughout this analysis. 

In principle a small map like those considered here will impose a high-pass filter that removes some of the CMB variance, and in practice using high-pass or matched filters are effective methods of cluster detection. However, low-$\ell$ CMB modes contribute very little to SZ contamination: the figures of merit (FoMs, ref. \S\ref{fom}) increase by $\sim0.1\%$ when we only consider $\ell\geq100$, $\sim1\%$ for $\ell\geq300$, and $\sim2\%$ for $\ell\geq500$. Thus we ignore any filter effects on CMB noise.

\subsection{Instrument noise}

We model the instrumental covariance between observations indexed by $\alpha$ and $\beta$ as
\begin{equation}
C_{\alpha\beta}^{\rm ins} = \frac{1}{T_\cmb^2} \frac{(T_a^{\rm sens})^2}{\Omega_{\rm pix}} \delta_{\alpha\beta},
\end{equation}
where $T_a^{\rm sens}$ is the beam sensitivity at $\nu_a$ corresponding to the index $\alpha$ (or $\beta$, because of the $\delta_{\alpha\beta}$), and $\Omega_\pix = \theta_\pix^2$ is the solid angle of the pixel. We assume that the correlated atmospheric noise is removed efficiently at these angular scales and do not include it in the analysis. 

Target map sensitivities and beam sizes are reported in Table~\ref{expt_param}, which describes both a baseline survey strategy and integration time, CCAT$_{base}$, as well as a future more sensitive instrument and survey strategy, CCAT$_{opt}$. For CMB experiments $T_a^{\rm sens}$ is typically reported in units of $\mu$K-arcmin, while radio and dusty galaxy observations are typically reported in Jy beam$^{-1}$, which can be converted to units of $\mu$K-arcmin using equation \ref{Jy_conv} and the relevant beam solid angle, $\Omega = \pi \xi^2/4\, {\rm ln}\,2$.

\subsection{Dusty Star Forming Galaxies}
\label{sec:dsfgs}

For upcoming surveys like CCAT-p the largest source of noise when separating SZ signals is expected to arise from dusty star forming galaxy (DSFG) emission. This is largely composed of radiation from dust at temperatures an order of magnitude greater than $T_\cmb$. These sources can significantly contaminate the SZ signal at higher frequencies. We take into account both the spatial randomness as well as spectral uncertainty on the variations of the dust temperature and redshift of the sources. 

If we expand the DSFG emission in spherical harmonics, we can express the observed covariance for this galactic component in a power series similar to the CMB covariance term,
\begin{equation}
\langle \Delta T_\gal^2 \rangle = \sum_\ell \frac{2\ell+1}{4\pi}C_\ell\exp \left (-\frac{\ell^2 \xi^2}{8\, {\rm ln}\, 2}\right ).
\end{equation}
We can estimate $C_\ell $ using the observed DSFG number counts $N(S)$ by expressing the sum over the sources as an integral over the flux $S$ of each source \citep[e.g.][]{1996MNRAS.281.1297T, 1999A&A...346....1S, knox}:
\begin{equation}
C_\ell  = \left ( \frac{\partial T^\cmb}{\partial B^\cmb_\nu} \right )^2 \int S^2 \frac{dN}{dS}\, dS.
\end{equation}

Since the galactic noise scales with the inverse of the beam solid angle $\langle\Delta T_\gal^2 \rangle \propto \Omega^{-1} \propto \xi^{-2}$ to leading order (because $C_\ell $ is constant), we define $\delta T_\gal \equiv \sqrt{\Omega} \Delta T_\gal$ to represent the true galactic noise, a property of the sky independent of the experiment. We obtain
\begin{equation}
\langle \delta T_\gal^2 \rangle = \frac{1}{2} \left ( \frac{\partial T^\cmb}{\partial B^\cmb_\nu} \right )^2 \int S^2 \frac{dN}{dS}\, dS.
\end{equation}

To estimate this quantity we use observations from the new Submillimeter Common-User Bolometer Array (SCUBA-2) Cosmology Legacy Survey (S2CLS) on the James Clerk Maxwell Telescope \citep{2013MNRAS.430.2513H}. \cite{scuba} fit point source counts at  850 $\mu$m (350 GHz) to the following form:
\begin{equation}\label{N(S)}
\frac{dN}{dS} = \frac{N_0}{S_0} \left ( \frac{S}{S_0} \right )^{-\gamma} \exp \left ( -\frac{S}{S_0} \right ),
\end{equation}
with best-fit values
\begin{align*}
N_0 &= 7180\, {\rm deg}^{-2} = 1.99\, {\rm arcmin}^{-2},  & \gamma &= 1.5,\\
S_0 &= 2.5\, {\rm mJy} \mapsto 98\, \mu \rm K\,arcmin^2.
\end{align*}
This gives
\begin{equation}\label{s2cls}
\left\langle\left (\delta T_{350}^\gal\right)^2 \right\rangle =\frac{\sqrt{\pi}}{4}N_0 S_0^2= (92\: \mu \rm K \, arcmin)^2,
\end{equation}
where $\delta T_{350}^\gal$ denotes the galactic noise at any point of the sky at 350 GHz. This is lower than the value used in \cite{knox} of 170 $\mu$K arcmin, which roughly agrees with the DSFG shot noise at 353 GHz reported in \cite{2014A&A...571A..30P}. A possible source of this discrepancy is the unresolved DSFG emission. For the majority of the forecasts presented here we use the S2CLS value in equation \ref{s2cls}. However, for the sake of completeness we also calculate all results with the higher DSFG noise level of 170 $\mu$K arcmin and compare these analyses to our results in \S \ref{results}. As expected, increasing the DSFG noise primarily increases the importance of foreground subtraction. 

To estimate $\delta T_\gal$ at any frequency, we find the spectral dependence of the galactic noise and use that to scale $\delta T_{350}^\gal$. Dusty emission can be reasonably modeled as an optically thin modified blackbody, given by
\begin{equation}\label{greybody}
S_\nu = S_0 \left (\nu(1+z)\right)^\beta B_\nu\left(\nu, \frac{T_{\rm dust}}{1+z}\right),
\end{equation}
where $S_\nu$ traces the flux at frequency $\nu$ from a dusty galaxy at redshift $z$, with dust at temperature $T_{\rm dust}$ and emissivity index $\beta$. 

\citet{ACT_galnoise} analyzed nine dusty galaxies using data from the Atacama Cosmology Telescope (ACT), fitting them to optically thick emission. They report that the observed $z$ and $T_{\rm dust}$ distributions are similar to those obtained from a $\beta=2$ fit to equation \ref{greybody}. The results are also consistent with \citet{SPT_galnoise}, where the authors analyzed 39 clusters using data from the South Pole Telescope (SPT) as well as the Atacama Large Millimeter Array (ALMA) and fit them to a $\beta=2$ model. We approximate the reported distributions to Gaussians with the following means and standard deviations, keeping $\beta$ fixed to $2$:
\begin{align*}
\langle z \rangle &= 4 & \sigma(z) &= 1\\
\langle T_{\rm dust} \rangle &= 43\,{\rm K} & \sigma(T_{\rm dust}) &= 8\,{\rm K}.
\end{align*}

We hence model the spectral variations of the spatial galactic noise uncertainties as
\begin{equation}\label{eq:dsfgnoise1}
f_4(\nu) = A \frac{\partial T^\cmb}{\partial B^\cmb_\nu} \left \langle S_\nu (\nu) \right \rangle,
\end{equation}
where the brackets denote an expectation value with respect to $z$ and $T_{\rm dust}$ at fixed $\beta=2$, and $A$ is a normalization factor chosen such that $f_4(350\,{\rm GHz})=1$. The spectral uncertainties themselves are just the standard deviation of the frequency distribution:
\begin{equation}\label{eq:dsfgnoise2}
f_5(\nu) = A \frac{\partial T^\cmb}{\partial B^\cmb_\nu} \sqrt{\left \langle S_\nu (\nu)^2 \right \rangle - \left \langle S_\nu (\nu) \right \rangle^2},
\end{equation}
where $A$ is the same normalization constant from equation \ref{eq:dsfgnoise1}.

The above analysis describes a single source, but for a number of sources following the same frequency distribution, the standard deviation will be smaller by $\sqrt{n}$, the square root of the number of sources in the observation. We estimate the number using the effective number density $N_{\rm eff}$ and beam solid angle $\Omega$, $n = N_{\rm eff}\Omega$. We find $N_{\rm eff}$ by averaging over the contributions to $\delta T_\gal^{350}$, using the model from \citet{scuba}, given in equation \ref{N(S)}:
\begin{equation}
N_{\rm eff} = \frac{\int S^2 \frac{dN}{dS}\int_S^\infty \frac{dN}{dS'} \,dS'\,dS}{\int S^2 \frac{dN}{dS}\, dS} = 0.96 \,\rm arcmin^{-2}.
\end{equation}

Thus we obtain the following expression for the galactic noise:
\begin{equation}\label{t_gal_eq}
\delta T_\gal (\nu, \pmb{\theta}) = \delta T_{350}^\gal \left (f_4(\nu) + (N_{\rm eff}\Omega)^{-\frac{1}{2}}f_5(\nu)\right) E(|\pmb{\theta}|, \xi)
\end{equation}
where $E(\theta, \xi)$ is a magnification factor that describes the effects of weak gravitational lensing of the dusty galaxies by the cluster itself. This is an important factor, especially for more massive clusters: \cite{2010MNRAS.406.2352L} estimate lensing contamination for typical clusters to be as significant a signal as the SZ effect. In practice a subset of clusters will also suffer from strong lensing contamination \citep[e.g.][]{2007MNRAS.376.1073Z, 2013ApJ...769L..31Z}. The effect of lensing is amplification of noise, and correlation between sky locations that would have otherwise been independent. Since the deflection angle is generally small compared to the beam and cluster sizes, we ignore this additional covariance and consider only the magnification factor. For these forecasts we assume $E(\theta, \xi)$ to be the step function given by \citet{knox}, which, based on \citet{1998MNRAS.297..502B}, returns a magnification factor ranging from 1 to 2.5 for points whose angular distance from the cluster center is within the beam size. In our forecasts, removing this lensing factor improves the FoMs (ref. \S \ref{fom}) by $\sim10\%$.

The total galactic noise covariance between observations $\alpha$, at frequency $\nu_a$ and pixel location $\pmb{\theta}_i$, and $\beta$, at $\nu_b$ and $\pmb{\theta}_j$, is then
\begin{equation}
\begin{split}
C^\gal_{\alpha\beta} &=\frac{1}{T_\cmb^2}\left \langle \left(\delta T_{350}^\gal\right)^2 \right \rangle c^{\rm BS}_{\alpha\beta} E_\alpha E_\beta\\ &\times \left (\frac{f_4(\nu_a) f_4(\nu_b)}{(\Omega_a+\Omega_b) /2}+ c^{\rm BS}_{\alpha\beta} \frac{f_5(\nu_a) f_5(\nu_b)}{N_{\rm eff}(\Omega_a+\Omega_b)^2 /4}  \right ),
\end{split}
\end{equation}
where $E_\alpha \equiv E(|\pmb{\theta}_i|,\xi_a)$ and $c^{\rm BS}$ is a shot noise covariance matrix between the pixels that includes beam smoothing, which is the only source of correlation between different pixels. In other words, $c^{\rm BS}$ is the covariance between observations with overlapping regions of the sky, which is a large factor when the pixel size is smaller than the beam size, normalized to unit covariance between identical pixel locations. It is given by
\begin{equation}
c^{\rm BS}_{\alpha\beta} = {\rm exp} \left ( -4\,{\rm ln}\, 2\, \frac{|\pmb{\theta}_i - \pmb{\theta}_j|^2}{ \xi_a^2+ \xi_b^2}\right ).
\end{equation}

\subsection{Other Potential Contaminants and Systematics}
\label{sec:othersources}

While other potential sources of SZ contamination exist, they are expected to be sub-dominant to the contributions described above. A primary goal of this paper is to study the optimal balance of sensitivities for extracting the SZ signals from the largest sources of contamination in individual clusters in order to help optimize the frequency balance in upcoming cluster surveys. We expect that any lower level contaminants will increase the need for sensitivity over a wider range of frequencies than the distribution studied here. This will only emphasize our conclusion that submillimeter measurements will become increasingly important for future cluster surveys. Here we briefly discuss potential contamination from radio sources, dusty galaxies within clusters, and emission from our galaxy, though we leave inclusion of these contaminants in forecasts to future work.

Luminous radio sources, such as active galactic nuclei (AGN), can contaminate SZ measurements at lower frequencies. Fortunately, the most luminous of these sources typically have power law spectral distributions, $S \propto \nu^\alpha$, that fall quickly with increasing frequency ($\alpha <0$), resulting in a small fraction of clusters with significant radio contamination. For example, \cite{lin2009} found that $<2$\% of clusters with $M_{200}>10^{14} M_{\odot}$ and $z=0.6$ are expected to have radio contamination with an amplitude approaching 20\% of the tSZ amplitude at 150 GHz. The number of contaminated clusters decreases both for higher masses and higher redshifts. At $z=0.6$ they also find that 1--7\% of clusters in the mass range $10^{14}$--$10^{15}$ $M_\odot$ will have contamination exceeding 5\% of the tSZ signal, which drops to 0.2--2\% at $z=1.1$. More recently, \cite{sayers/etal:2013} found that only about 1/4 of the massive clusters in their sample showed a fractional change in the 140 GHz tSZ signal larger than 1\%. This contamination will also be reduced at the higher frequencies being measured with CCAT-p. 

Radio sources that would otherwise contaminate SZ measurements can often be found in existing radio catalogs as described in \cite{lin2009} and \cite{sayers/etal:2013}. This enables simple removal of several percent of the clusters in the catalog due to radio contamination. As long as the clusters with luminous AGN or other sources are removed from the SZ extraction catalog, separate radio noise terms can be ignored in the analysis presented here. In practice, one can also subtract known radio sources from contaminated clusters, but this process still results in loss of information, and is not considered here because so few clusters appear to be contaminated by radio sources.

Dust emission from the galaxy clusters themselves is another expected source of contamination. This has the potential to bias the SZ effect as it also traces the cluster profile, but can be distinguished spectrally; once again emphasizing the need for future experiments to be sensitive to submillimeter wavelengths. Evidence for cluster dust was recently presented in \cite{planck_szdust:2016}. It is a weak effect: hundreds of cluster measurements were stacked to extract these signals and compare them to infrared measurements. A weak correlation between the CIB and optical galaxy clusters was also reported in \cite{2013JCAP...05..004H}. \citet{erler}, who measure the dusty emission that is correlated with the hundreds of clusters in their sample, report an amplitude of 8~mJy\,arcmin$^{-2}$ at 857~GHz, which is $\sim2.5\times$ smaller than the total amplitude of DSFG noise considered here (ref. Figure~\ref{fig:frequency_dist}). Thus, while this dust emission is not included in our current analysis, it will be important to understand for accurate cluster constraints in the future, and  CCAT-p submillimeter measurements are expected to be valuable for characterizing this component. 

In addition, dust within our own galaxy can contaminate the SZ signals. This dust varies strongly with position on the sky, and is not expected to be a dominant contaminant for low-dust fields at high galactic latitudes that only cover roughly $10^3$ deg$^2$ \citep{planck_foregrounds:2015}.

Among possible systematic effects, we observe that the assumption of isothermality for the intracluster medium often does not reflect the properties of real clusters \citep{0004-637X-567-1-163,0004-637X-725-2-1452}. \cite{lindner} applied the peculiar velocity measurement method to SZ simulations and found that assuming isothermal clusters could introduce a bias between the real and recovered velocity ranging from $15\%$ to $36\%$ depending on the aperture used to measure the SZ effects. Other effects of non-isothermality are briefly described in \S\ref{ind_res}.

In this analysis we are neglecting internal flows of the electron gas, which can be as large as the overall peculiar velocity of the cluster \citep{2001ApJ...551..160V,2002ApJ...569L..31M,2002ApJ...576..708S,0004-637X-587-2-524,2004ApJ...606..819M,2017A&A...598A.115A}. \cite{0004-637X-587-2-524} have verified that internal flows introduce a dispersion in the peculiar velocity estimated from the kSZ of $50-100$ km/s. This effect is small but not negligible, especially for sensitive experiments approaching 1\,$\mu$K-arcmin map noise levels, and may represent a lower limit on the precision of kSZ estimated peculiar velocities, unless internal flows can be accounted for. 

We also assume the cluster to be perfectly centered in the map. In principle one can always generate a perfectly centered map from the time stream data, but the effect of miscentering in individual clusters is small. Pixel-scale offsets cause the FoMs (ref. \S \ref{fom}) to change by less than $1\%$, while offsets on the scale of the cluster size $\tfoo$ cause the FoMs to decrease by $\sim2\%$. These effects are related to the variation of the forecast results with map and pixel sizes, which is described in Appendix \ref{sec:pixel}. The effect of miscentering is expected to be larger for differential analyses and for cross-correlation analyses involving multiple clusters, and is investigated in \cite{2017arXiv171001755C}, but in these cases it is the relative miscentering between clusters that is important.

The assumption of a specific pressure profile can potentially be a source of bias for the results, although measurements of both the tSZ and kSZ effects have shown that the observed signals have a relatively weak dependence on the details of the assumed profile, for example when using a matched filter approach to remove the CMB and noise \citep{2013JCAP...07..008H,2016MNRAS.461.3172S}. The assumption of a specific temperature profile may be a more important source of bias. Regardless, if X-ray observations are available, these biases and uncertainties can be reduced further.

\section{Dusty Galaxy Subtraction}\label{foreground}

For experiments equipped with submillimeter wavelength detectors, the emission from DSFGs may be partially subtracted by scaling the high frequency maps by the expected DSFG scaling relationships and subtracting them from the lower frequency maps to reduce galactic noise. This follows the approach in \cite{lindner}.

However, because the spectral dependence of the galactic noise varies spatially, there will still be errors from this imperfect subtraction, and the instrumental noise of the higher frequency bands will affect every band. The SZ signal and CMB noise are also reduced by their scaling factors. However, since DSFG contamination is such a large source of noise, experiments with sensitive submillimeter wavelenths will be able to improve the measurements of SZ parameters with this approach.

If we have $n_{\rm fg}$ foreground bands at frequencies $\nu_f$, 
then simply averaging them and subtracting the average from the regular bands leads to the corrected temperature,
\begin{equation}\label{fgsubgen}
\begin{split}
    \Delta T'(\nu, \pmb{\theta}) &= \Delta T(\nu, \pmb{\theta}) - \frac{1}{n_{\rm fg}}\sum_f\frac{f_4(\nu)}{f_4(\nu_f)}\Delta T(\nu_f, \pmb{\theta})\\&\equiv\Delta T(\nu, \pmb{\theta}) - \mean{\frac{f_4(\nu)}{f_4(\nu_f)}\Delta T(\nu_f, \pmb{\theta})}.
\end{split}
\end{equation}
The foreground maps are typically more resolved, since they use observations that are usually at higher frequencies. In any case, the more resolved maps are assumed to be smoothed to match the least resolved one, so that we may ignore effects due to differing beam sizes. Since the second term of equation \ref{fgsubgen} only differs in frequency from the first, we can absorb the foreground subtractions into the frequency dependencies $f_i \rightarrow f'_i$:
\begin{equation}\label{fprime}
f'_i(\nu) = f_i(\nu) - \mean{\frac{f_4(\nu)}{f_4(\nu_f)}f_i(\nu_f)}
\end{equation}
for $i\leq 4$, while for $f'_5$ we have
\begin{equation}\label{f5prime}
f'_5(\nu) = A\frac{\partial T^\cmb}{\partial B^\cmb_\nu}\sqrt{\left \langle \left( S_\nu(\nu) - \mean{\frac{\langle S_\nu(\nu)\rangle}{\langle S_\nu(\nu_f)\rangle} S_\nu(\nu_f)} \right)^2 \right \rangle},
\end{equation}
where $A$ is the same normalization constant from equations \ref{eq:dsfgnoise1} and \ref{eq:dsfgnoise2}, i.e. it sets $f_4(350\,{\rm GHz})=1$ before subtraction.

Since $f'_4=0$, the spatial term of the galactic noise is eliminated, leaving only the spectral term proportional to $f'_5$:
\begin{equation}
\delta T_\gal (\nu, \pmb{\theta}) = \delta T_{350}^\gal (N_{\rm eff}\Omega)^{-\frac{1}{2}}f'_5(\nu) E(|\pmb{\theta}|, \xi).
\end{equation}
The galactic noise covariance matrix reduces to
\begin{equation}\label{X}
C^\gal_{\alpha \nu} =\frac{1}{T_\cmb^2}\left \langle \left(\delta T_{350}^\gal\right)^2 \right \rangle \frac{\left(c^{\rm BS}_{\alpha\nu}\right)^2 E_\alpha E_\nu f'_5(\nu_a) f'_5(\nu_b)}{N_{\rm eff}(\Omega_a+\Omega_b)^2 /4}.
\end{equation}

The CMB term, originally independent of frequency, now has the dependence
\begin{equation}
f'_\cmb(\nu) = 1 - \mean{\left(\frac{f_4(\nu)}{f_4(\nu_f)}\right)},
\end{equation}
and the corresponding covariance thus has the following frequency dependence
\begin{equation}\label{subcmb}
C^\cmb_{\alpha\beta} = C^\cmb_{\alpha\beta} f'_\cmb(\nu_a) f'_\cmb(\nu_b).
\end{equation}
These factors are smallest when $\nu_f \simeq \nu$, which is when the scaling is close to unity. However, to prevent elimination of the signal, foreground frequencies should not be too close to signal frequencies.

Since the foreground map is subtracted from other channels, the instrument noise of the foreground map is present in the other maps. The instrument noise of frequencies with greater angular resolutions (smaller beam sizes) would be suppressed by smoothing the maps to match the lowest angular resolution, but we ignore this factor to maintain more conservative forecasts. The instrumental noise covariance term then becomes
\begin{equation}\label{subins}
 C_{\alpha\beta}^{\rm ins} = \frac{1}{T_\cmb^2} \frac{1}{\Omega_{\rm pix}}\left ( (T_a^{\rm sens})^2 + \frac{1}{n_{\rm fg}}\mean{\left(\frac{f_4(\nu_a)}{f_4(\nu_f)}T_f^{\rm sens}\right)^2} \right )\delta_{\alpha\beta}
\end{equation}
The final covariance matrix for foreground subtraction is hence given by the sum of equations \ref{X}, \ref{subcmb} and \ref{subins}.

\section{Experiment parameters} \label{experiments} 

CCAT-p is a 6-meter aperture telescope that will be built near the top of Cerro Chajnantor at 5600 meters elevation in the Atacama desert in Chile with first light planned for 2021. The telescope will have a large field-of-view (FOV), roughly 8$^\circ$ diameter, and a half wavefront error near 10 microns. The high-throughput optics will enable illumination of much larger detector arrays than previous millimeter and submillimeter telescopes \citep{niemack:2016}. 

Here we study a potential first light instrument configuration for CCAT-p that would utilize roughly 1/7 of the available FOV. The concept is based on the instrument described in \cite{stacey/etal:2014}, with seven separate optical paths; however, the wavelengths assigned to each optical path are considerably different here. In addition to a broader wavelength range, at frequencies below 500~GHz we assume the use of dichroic detector arrays similar to those described in \cite{datta/etal:2016}. Single frequency detector arrays are assumed at higher frequencies where dichroic arrays have not yet been demonstrated. The bands have been selected to match the telluric windows accessible from the CCAT-p site. A 862~GHz band is desired for both studying star formation history in sub-millimeter galaxies and characterizing dust emission from clusters \citep{erler}.

\begin{deluxetable}{cc|cc|cc}
\tablecaption{Experimental parameters for the ``non-optimized'' CCAT$_{base}$ baseline configuration with seven optical paths and the CCAT$_{opt}$ optimized configuration with about thirty optical paths, \S\ref{sec:pair_det}.
\label{expt_param}}
\tablehead{\colhead{$\nu$} & \colhead{$\xi$} & \colhead{$A_{\rm base}$} & \colhead{$T^{\rm sens}(\rm base)$} & \colhead{$A_{\rm opt}$} & \colhead{$T^{\rm sens}(\rm opt)$} \\
GHz & arcmin & arrays & $\mu$K-arcmin\tablenotemark{b} & arrays & $\mu$K-arcmin\tablenotemark{c}}
\startdata
95 & 2.2 & 2 & 4.9 & 16 & 0.9\\
 150 & 1.4 & 2  & 6.4 & 15 & 1.2\\
  226 & 1.0 & 2  & 4.9 & 3 & 2.0\\
  273 & 0.8 & 2  & 6.2 & 5 & 2.0\\
  350\tablenotemark{a} & 0.6 & 2  & 25 & 4 & 8.9\\
  405\tablenotemark{a} & 0.5 & 2  & 72 & 5 & 23\\
  862 & 0.3 & 1 & $6.6\times 10^4$ & 2 & $2.3\times10^4$

\enddata
\tablenotetext{a}{Bands used only to subtract foreground galactic noise emission, unless otherwise specified}
\tablenotetext{b}{Map sensitivities assume 4000 hours observing a 10$^3$ deg$^2$ field.}
\tablenotetext{c}{Map sensitivities assume 16,000 hours observing a 10$^3$ deg$^2$ field, or deployment of more arrays and a shorter time.}
\end{deluxetable}

Table \ref{expt_param} shows map sensitivity estimates for a 4000~hour, $10^3\,\rm{deg}^2$ survey. This configuration, which we refer to as the CCAT$_{base}$ configuration, would enable CCAT-p to simultaneously observe from 95~GHz up to 862~GHz in seven frequency bands, overlapping and complementing the frequency coverage of current CMB surveys. We consider four possible survey strategies in terms of observing time and survey areas, given by the combinations \{$4\times10^3$~hr, $1.2\times10^4$~hr\}$\times$\{$10^3\,\rm{deg}^2$, $10^4\,\rm{deg}^2$\}.

We also consider how CCAT-p measurements could be improved through three different approaches: 1) optimizing the balance of CCAT$_{base}$ frequencies to extract the SZ signals; 2) combining CCAT-p data in the configuration described above with planned CMB measurements from Advanced ACTPol \citep[AdvACT,][]{Henderson:2015nzj} on the Atacama Cosmology Telescope \citep{fowler/etal:2007}; and 3) an optimized future upgrade to CCAT-p, which we call CCAT$_{opt}$, filling most of the available FOV with about 30 separate optical paths using 16,000 hr of observations. This instrument would provide some of the desired capabilities for a Stage-IV CMB survey \citep[CMB-S4, e.g.,][]{abazajian/etal:2015, abazajian/etal:2016}, though greater sensitivity via multiple telescopes will be needed to accomplish the full CMB-S4 science goals.

\section{Results}\label{results}

We first present forecast results for individual clusters, then for distributions of clusters detected in large area surveys. Finally, we present the frequency band optimization in order to study different instrument configurations and assess the value of submillimeter bands for these measurements.

After trying all possible combinations of foreground bands, with equal weights and with weights chosen to minimize total instrument noise, we found that the equal weighting of the 350 and 405 GHz channels achieves the best FoM results for a wide variety of clusters, so we use that configuration for the forecasts presented here.

\begin{figure*}
\begin{center}
\large{CCAT$_{base}$}
\end{center}
\includegraphics[width=.5\textwidth]{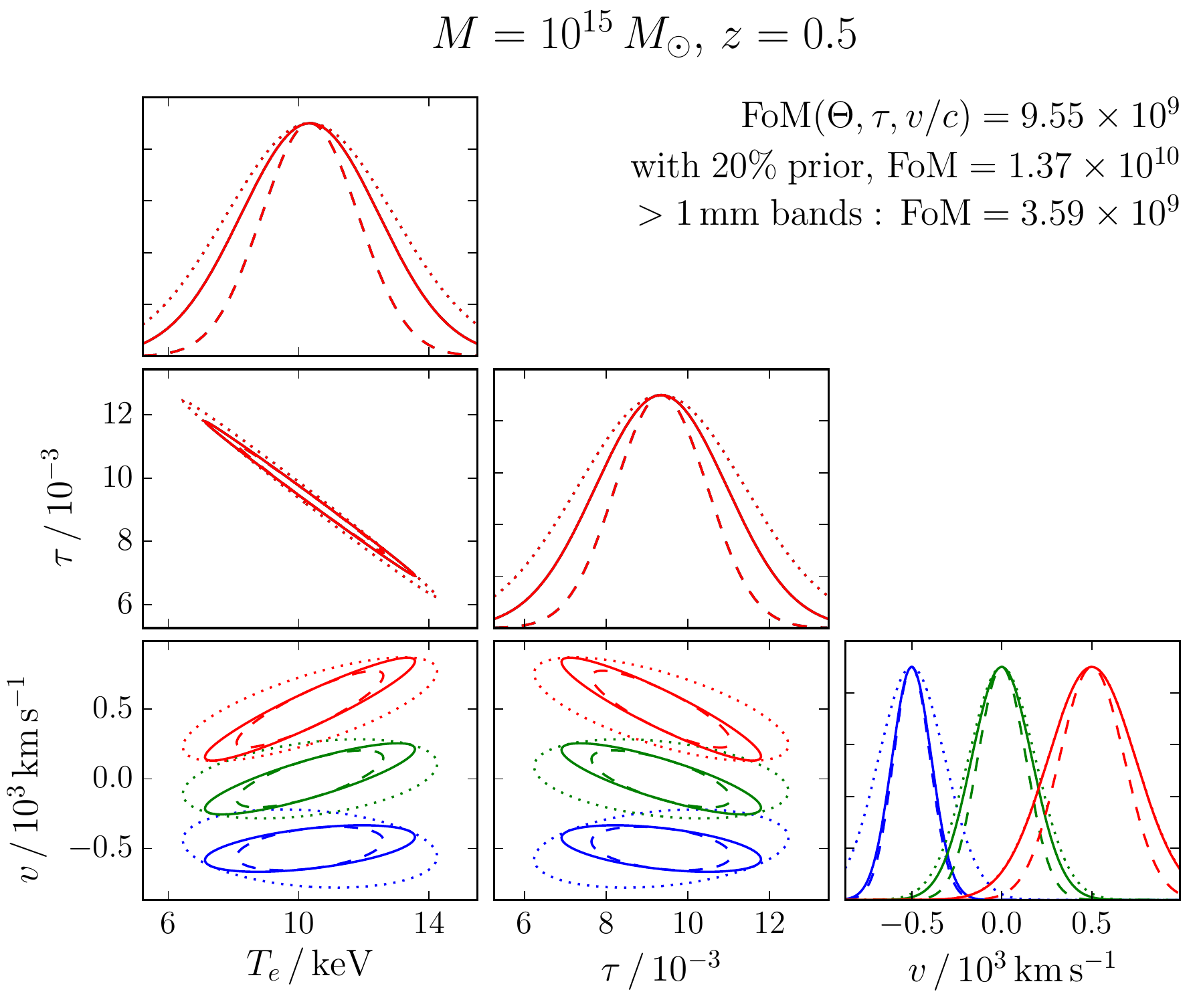}
\includegraphics[width=.5\textwidth]{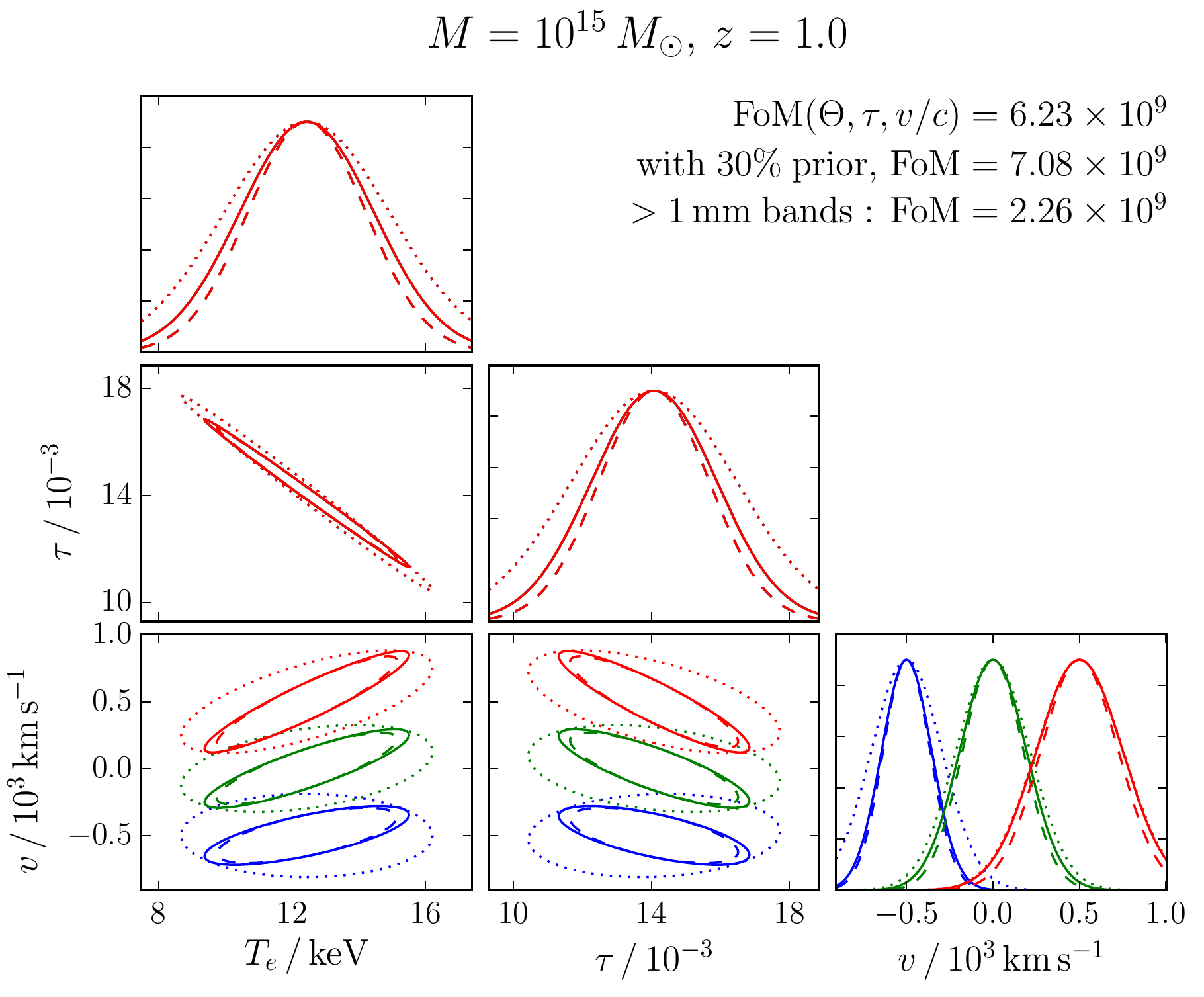}
\includegraphics[width=.5\textwidth]{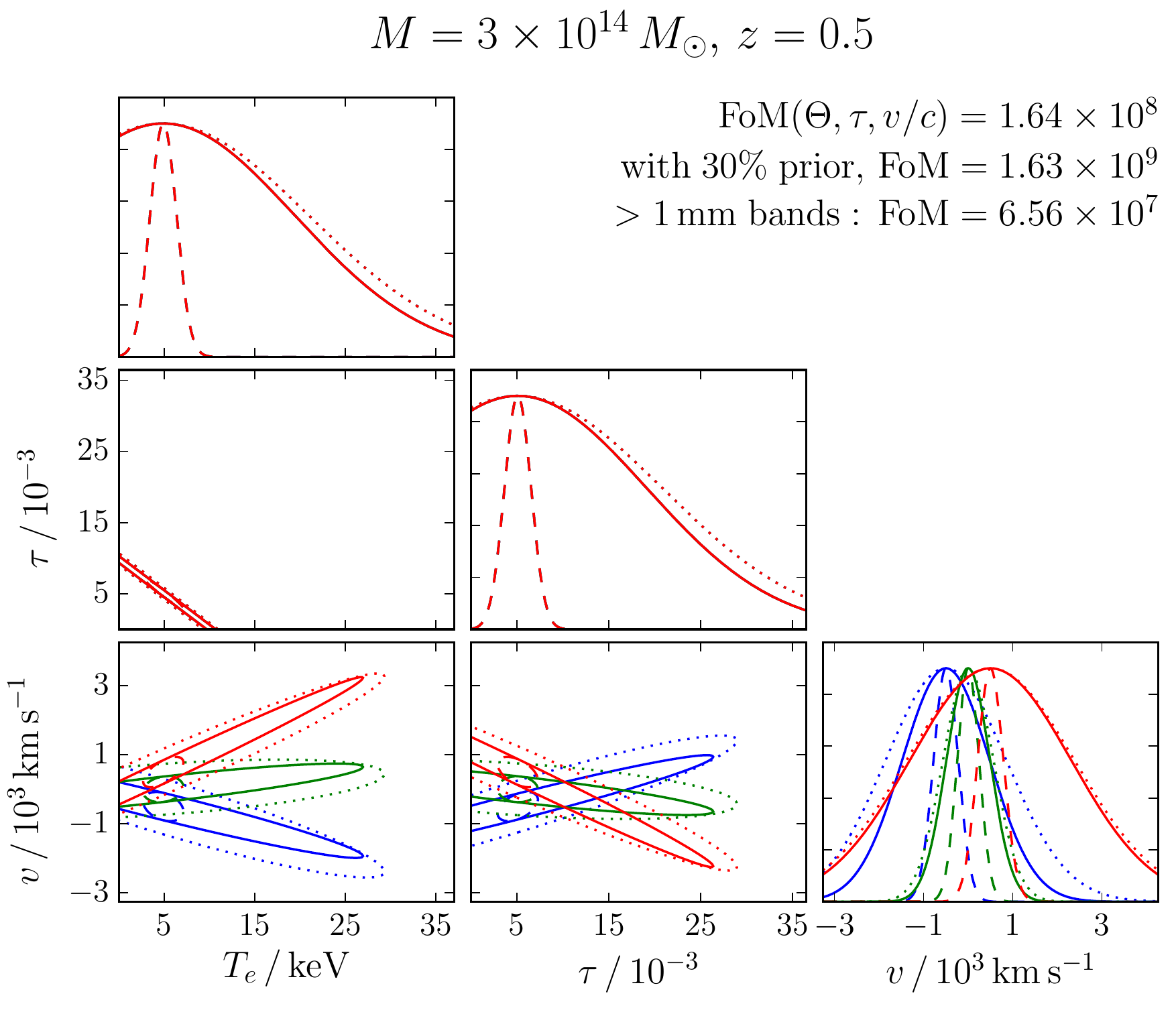}
\includegraphics[width=.5\textwidth]{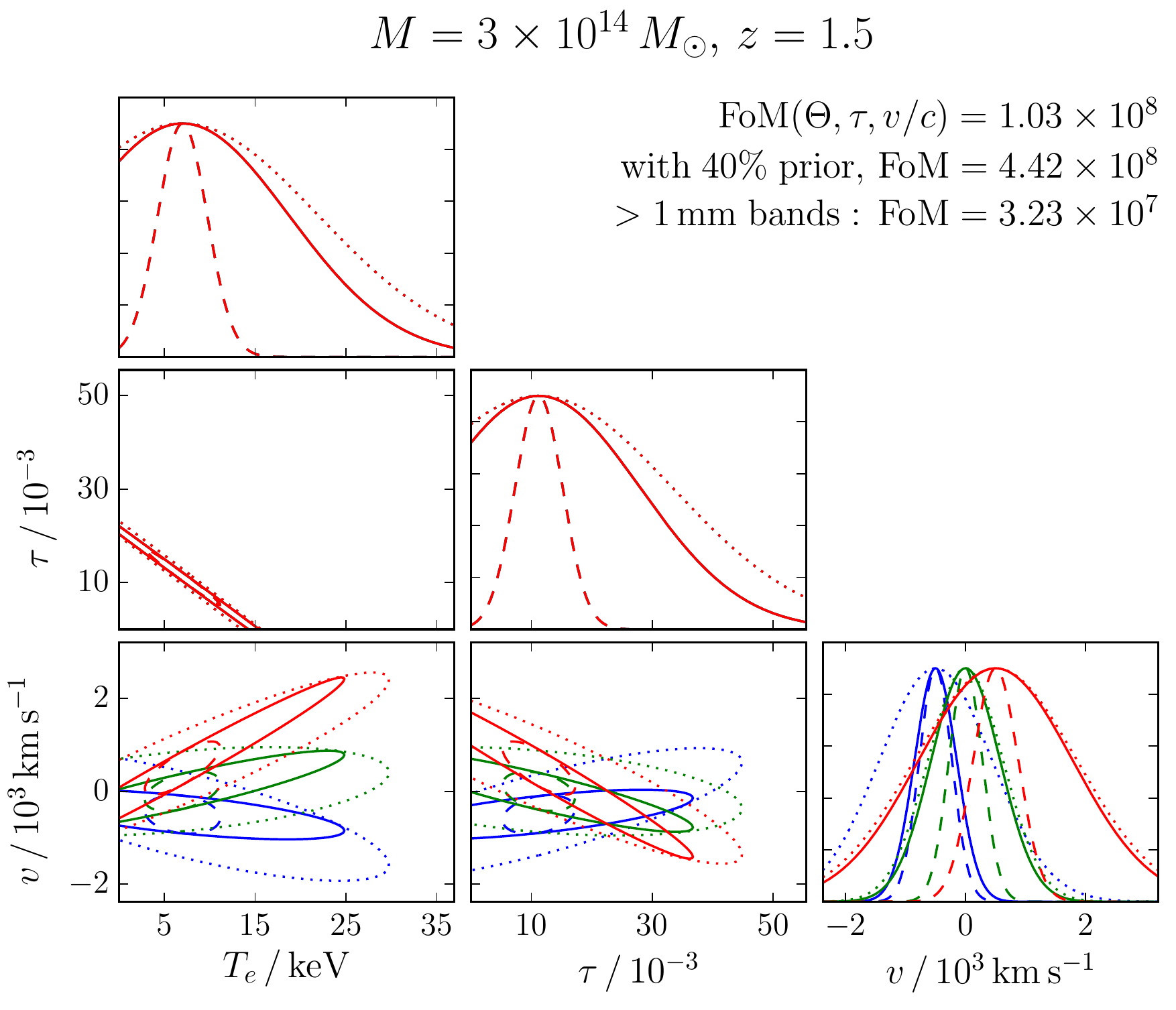}
\caption{Expected 1-$\sigma$ confidence contours from CCAT-p for the cluster parameters, $\tau$, $T_e$ and $v$, for fiducial clusters of various masses and redshifts. We have plotted three fiducial velocities for each cluster, represented by the different colours. As can be seen from the peaks of each velocity distribution, these fiducial velocities are $-$500, 0, and 500 km s$^{-1}$. All of the plots are for the CCAT$_{base}$ survey, as described in Table~\ref{expt_param}. The dotted lines depict the constraints when only $>1$ mm bands are used, while the dashed lines show the constraints including a prior on $T_e$ from eROSITA. Based on \citet{Borm:2014zna}, we use a 20\%, 30\%, or 40\% prior on the electron temperature as indicated in each quadrant, which may be an optimistic interpretation of \citet{Borm:2014zna} for the higher $z$ clusters. The FoM is independent of velocity, because the Fisher matrix in $P_1,P_2,P_3$ is independent of velocity, and so is the determinant of the transformation matrix (ref. equation~\ref{fom_trans}). Furthermore, covariances involving $T_e$ and $\tau$ are also independent of velocity as a result of equation~\ref{cov_trans_matrix}. When the level of DSFG noise is increased (see \S\ref{sec:dsfgs}), the FoMs decrease by $\sim 10 \%$ including submillimeter bands and decrease by 20--30\% without submillimeter bands.
\label{contour_op1}}
\end{figure*}

\begin{figure*}
\begin{center}
\large{CCAT$_{opt}$}
\end{center}
\includegraphics[width=.5\textwidth]{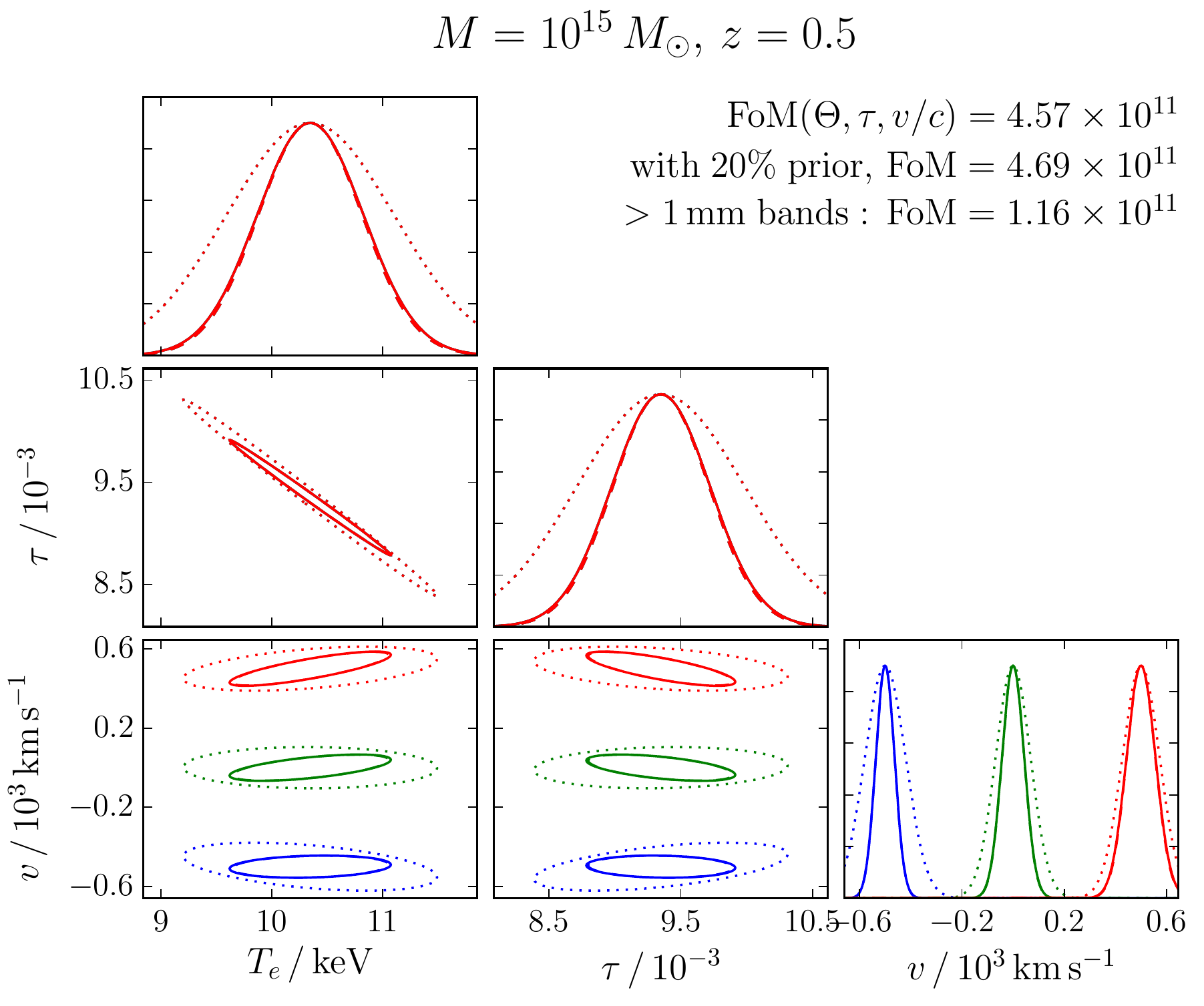}
\includegraphics[width=.5\textwidth]{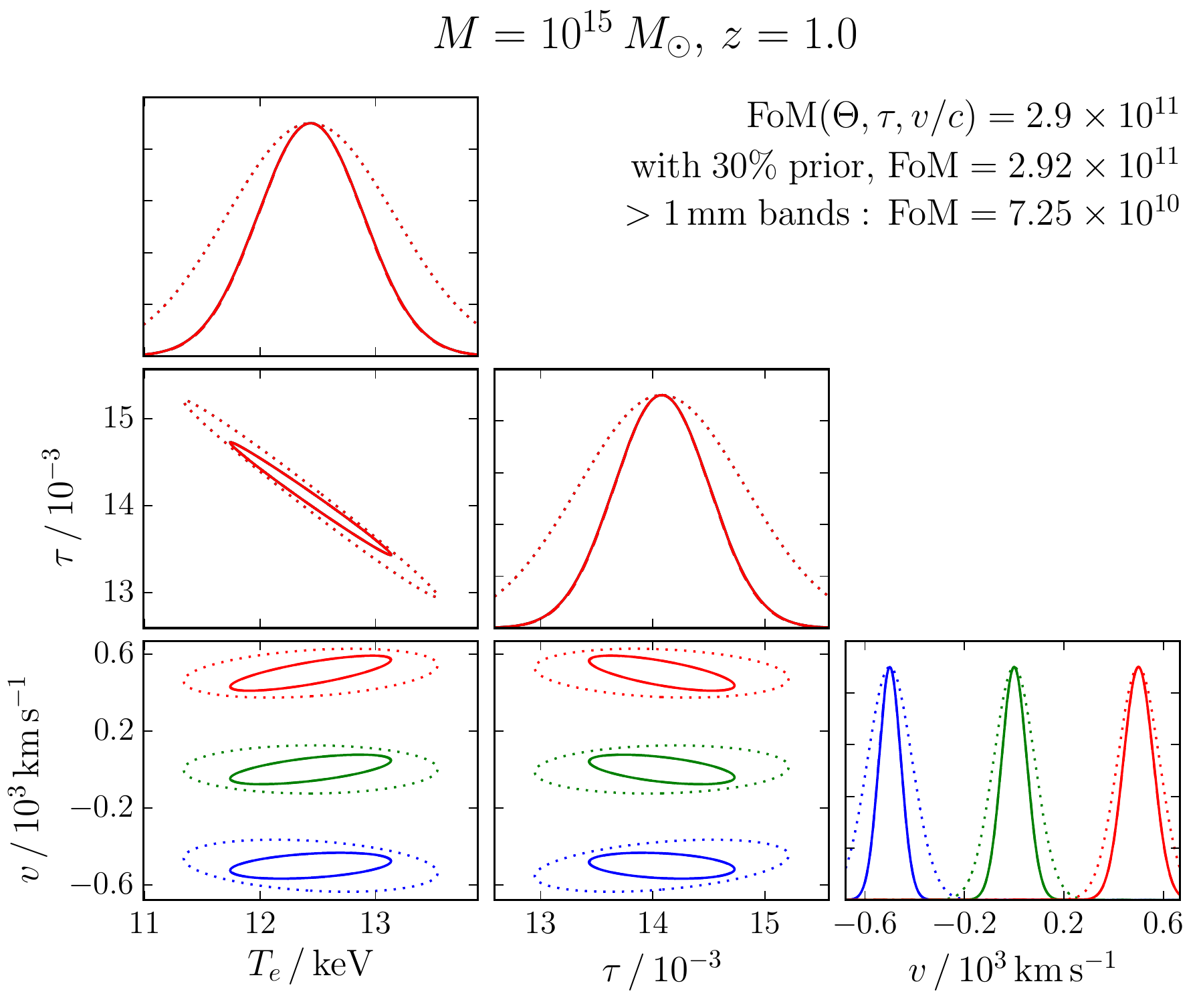}
\includegraphics[width=.5\textwidth]{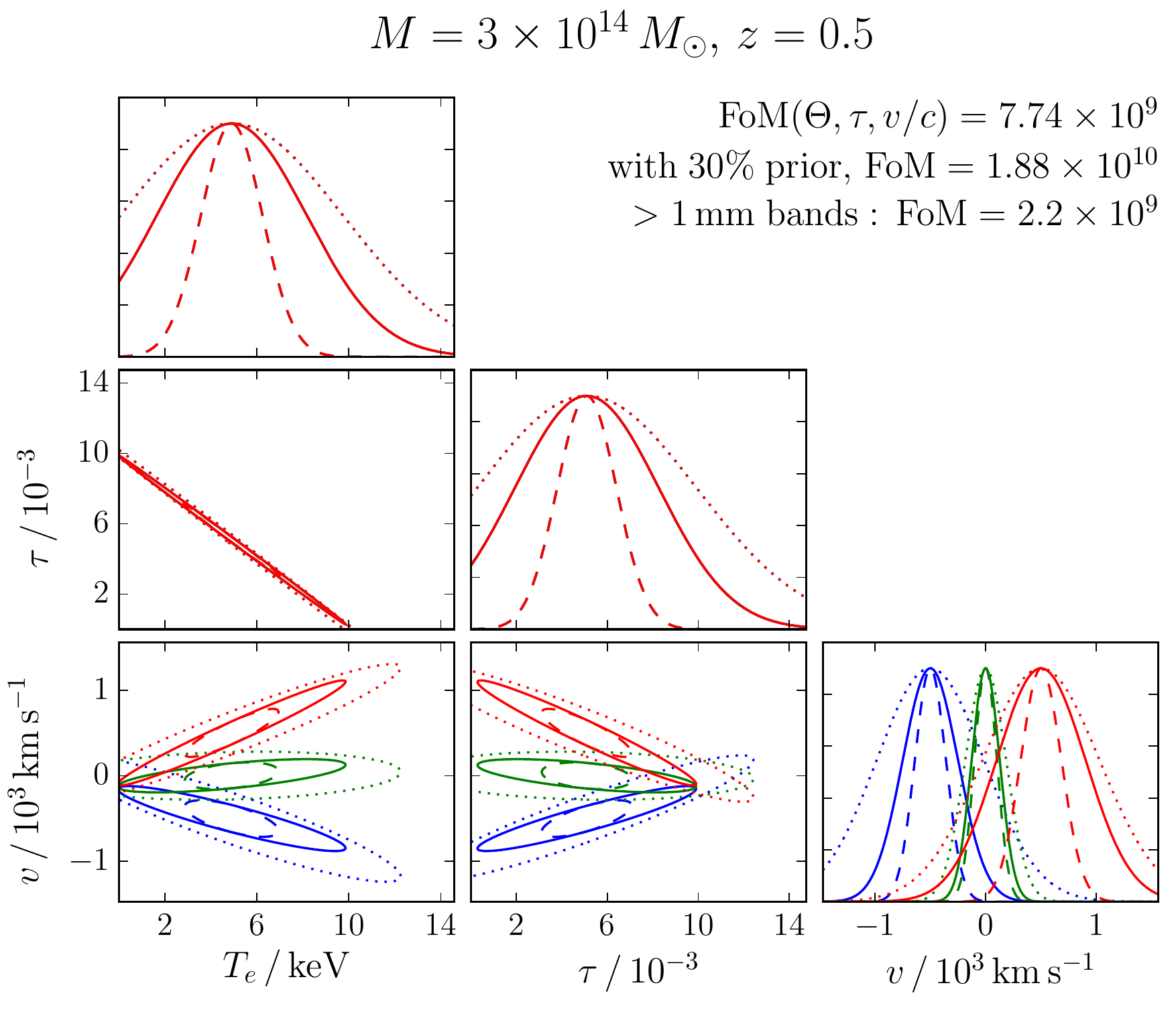}
\includegraphics[width=.5\textwidth]{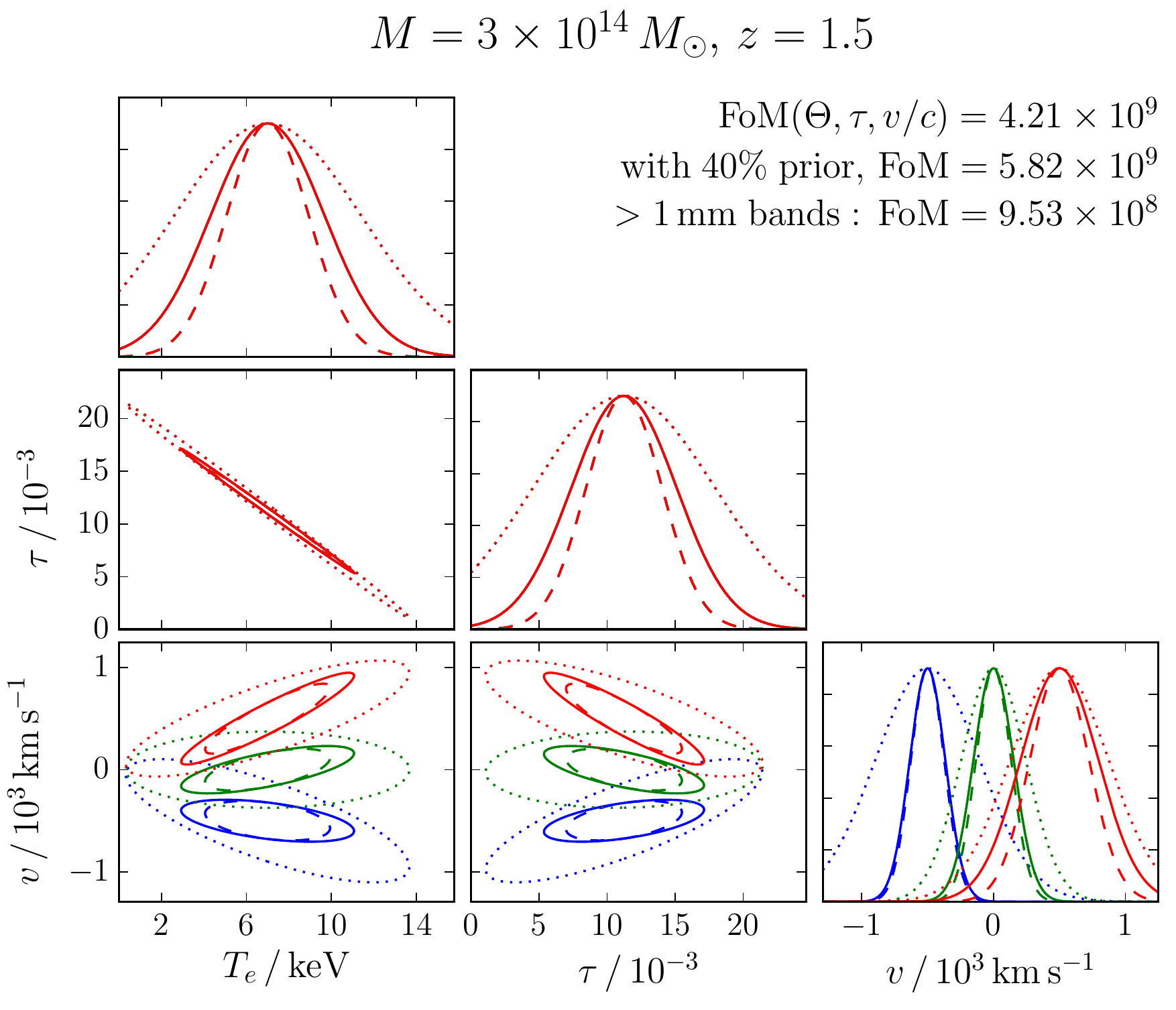}
\caption{Same as Figure \ref{contour_op1}, for CCAT$_{opt}$ sensitivities indicated in Table~\ref{expt_param}. Note that the CCAT$_{opt}$ measurements alone provide strong constraints on $T_e$ and other parameters for high mass clusters (top panels). Even at lower masses (bottom panels), the constraints are dramatically better than those of CCAT$_{base}$. When the level of DSFG noise is increased (see \S\ref{sec:dsfgs}), the FoMs decrease by $\sim 30 \%$ including submillimeter bands and decrease by 30--40\% without submillimeter bands. Further comparison with Figure \ref{contour_op1}  shows the submillimeter bands make a larger difference for CCAT$_{opt}$, indicating the increased importance of a broader frequency range as measurements improve. 
\label{contour_op4}}
\end{figure*}

\subsection{Individual cluster results}\label{ind_res}

\citet{knox} forecast uncertainties for a fiducial cluster. In an attempt to pick more realistic cluster parameters, we assume a mass $M\equiv M_{500}$ and redshift $z$ for each cluster, and use best-fit scaling relations to transform them into the SZ parameters $T_e$ and $\tau$. We use the $T$-$M$ scaling relation from \citet{2016MNRAS.463.3582M}:
\begin{equation}\label{TM}
\frac{T_e}{\rm keV} = 8.76 \times \left ( E(z) \frac{M_{500}}{10^{15}M_\odot}\right )^{0.62},
\end{equation}
where $E(z)\equiv H(z)/H_0$. For the $\tau$ scaling, we use the $Y$-$M$ scaling from \cite{Ade:2013lmv}:
\begin{equation}\label{YM}
Y_{500} = 10^{-4.19}  E(z)^{\frac{2}{3}} \left (\frac{d_A(z)}{\rm Mpc}\right )^{-2} \left ( \frac{M_{500}}{7.5\times 10^{14}M_\odot}\right )^{1.79},
\end{equation}
where $d_A(z)$ is the angular diameter distance and $Y_{500}$ is the integrated comptonization parameter within $\tfoo$. Using the cluster model from \S\ref{SZ} and Figure \ref{int_signal_plot}, we write
\begin{equation}\label{int_y_eq}
\frac{Y(\theta')}{y} = 2\pi \int_0^{\theta'} \theta h(\theta)\,d\theta,
\end{equation}
where $y \propto \tau T_e$. We find that $Y_{500} = 0.423 \,y\, \tfoo^2$. The cluster size $\tfoo$ is simply related to $d_A(z)$ and $R_{500}$, which is related to $M_{500}$ by the critical density of the universe $\rho_{\rm crit}(z)$. Thus, assuming $M,z,v$ for a cluster completely determines all relevant cluster parameters. Since the scaling relations involve significant scatter, we use these only to obtain more physically relevant fiducial cluster parameters, and do not change basis from $T_e,\tau,v$ to $M,z,v$. 

Figures \ref{contour_op1} and \ref{contour_op4} show the expected 1-$\sigma$ contours for $\tau$, $T_e$ and $v$ for a few combinations of cluster parameters measurable by CCAT-p (cluster masses $3\times10^{14}M_{\odot}$ and $10^{15}M_{\odot}$, redshifts $0.5$, $1$, and $1.5$) for the two surveys, CCAT$_{base}$ (Figure \ref{contour_op1}) and CCAT$_{opt}$ (Figure \ref{contour_op4}), described in Table~\ref{expt_param}.
To compare with more traditional CMB surveys, we also plot the constraints achieved by using data from wavelengths $>1$ mm ($\nu<300$ GHz) only. We find that CCAT-p does far better than current sub-millimeter telescopes: CCAT$_{base}$ FoMs are $\sim15\times$ greater than those of AdvACT, and the FoMs of CCAT$_{opt}$ are almost $50\times$ greater than those of CCAT$_{base}$. These figures also confirm that the degeneracies between these three parameters are significant and represent one of the main obstacles to separating optical depth, temperature, and velocity.

External X-ray data can provide independent constraints on the temperature $T_e$ and thereby help in breaking some of the degeneracies. For example the eROSITA satellite \citep{2010SPIE.7732E..0UP,2012arXiv1209.3114M}  is expected to provide a high spectral and angular resolution full-sky survey in the medium energy X-ray range. \cite{Borm:2014zna} found that eROSITA can constrain cluster temperatures with a $10\%$ precision for clusters up to redshift $z\simeq0.16$. At higher redshifts up to $z\simeq1$ or possibly $1.5$, and depending on the cluster mass, eROSITA measurements are expected to constrain the temperature with a relative uncertainty ranging from $10$\% to $40\%$. 

To illustrate the potential of X-ray temperature constraints, we include priors on $T_e$ loosely based on eROSITA forecasts for an exposure time of 1,600 s \citep{Borm:2014zna}. The dashed lines in Figures \ref{contour_op1} and \ref{contour_op4} show the results when this prior is included in the calculation. Since the tSZ amplitude $P_1$ (ref. equation \ref{p1p2p3}) is the dominant signal and thus the most well-constrained parameter, the strongest degeneracy is between $T_e$ and $\tau$, so that a prior on the temperature almost directly translates into a prior on the optical depth. We note that X-ray measurements also provide information about the electron number density, which could be used to improve constraints on $\tau$ further or help quantify non-isothermal effects; however, for this analysis we only include priors on $T_e$. Depending on the cluster mass and peculiar velocity value, the temperature prior can reduce the uncertainty on the peculiar velocity by up to a factor of 3 in the fiducial clusters studied here.

On the other hand, when considering a much more sensitive survey, such as CCAT$_{opt}$ with CMB-S4 scale sensitivity, we find that in $10^{15} M_{\odot}$ clusters $T_e$ is well determined by the microwave measurements (Figure \ref{contour_op4} top) with negligible improvement from the X-ray prior. It is important to note that while X-ray measurements constrain the emission-weighted temperature, the SZ effect is determined by the temperature weighted by optical depth (ref. \S\ref{SZ}). Under our assumption of isothermality, these two are the same, but in practice using an X-ray prior in this way could be a potential source of bias. Regardless, X-ray observations will help in characterizing real clusters that are not spherical or isothermal, providing information about cluster profiles that will be useful in SZ measurements, which highlights one area of great complementarity  between future X-ray and microwave measurements.

Relaxing the assumption of isothermality but using the fixed power law temperature profile from \citet{2008A&A...486..359L} discussed in \S\ref{SZ}, we find kSZ uncertainties $\sigma(P_3)$ decrease and rSZ uncertainties $\sigma(P_2)$ increase by 30--40\%. It also causes tSZ uncertainties $\sigma(P_1)$  to improve by 5--10\%. Thus the FoM change is within 10\%, with smaller clusters typically gaining more than larger ones, and with a larger effect on CCAT$_{base}$ than CCAT$_{opt}$. The velocity uncertainties $\sigma(v)$ increase by 0--20\% for CCAT$_{base}$ (with smaller clusters affected more than larger ones) and decrease by $\sim 5\%$ for CCAT$_{opt}$. Since the effects of non-isothermality are small and can be constrained e.g. by X-ray observations, we do not consider non-isothermality for the following results.

\subsection{Cluster distribution results}

Building on the forecasts for individual clusters, we study constraints for different distributions of clusters probed via different survey strategies. While the baseline strategy, CCAT$_{base}$, assumes a survey length of 4,000 hr and area of 1,000 deg$^2$, we forecast results for the same instrument with both longer duration 12,000 hr surveys and with the area increased to 10,000 deg$^2$. We also forecast the results of a next generation upgraded survey (CCAT$_{opt}$), with roughly 30 optics tubes instead of 7, as described in Table~\ref{expt_param}, and a survey time of 16,000 hr over 1,000 deg$^2$. 

We first estimate the number of detected clusters. To obtain a minimum mass detection criterion, we compare the map-noise for the band most sensitive to the dominant tSZ signal (i.e. that with least $T^{\rm sens}/|f_1(\nu)|$) as a function of cluster mass (equation \ref{YM}). In principle one could take advantage of multi-frequency channels; however, the single band approach has been successfully used by several collaborations \citep[e.g.][]{Hasselfield:2013wf}. For simplicity, we base our estimates on the most sensitive band only. 
Specifically, we use the 95 GHz sensitivity of $4.9\,\mu \rm K\,arcmin$ and $0.9\,\mu \rm K\,arcmin$ for the two strategies described in Table \ref{expt_param}. We note that the 273~GHz CCAT$_{base}$ channel is expected to have comparable (or slightly better) sensitivity to the tSZ signal at $z>1$; however, we assume detection rates based on 95 GHz decrement measurements, which are known to agree well with recent tSZ detections.

\begin{figure*}
\includegraphics[width=.5\textwidth]{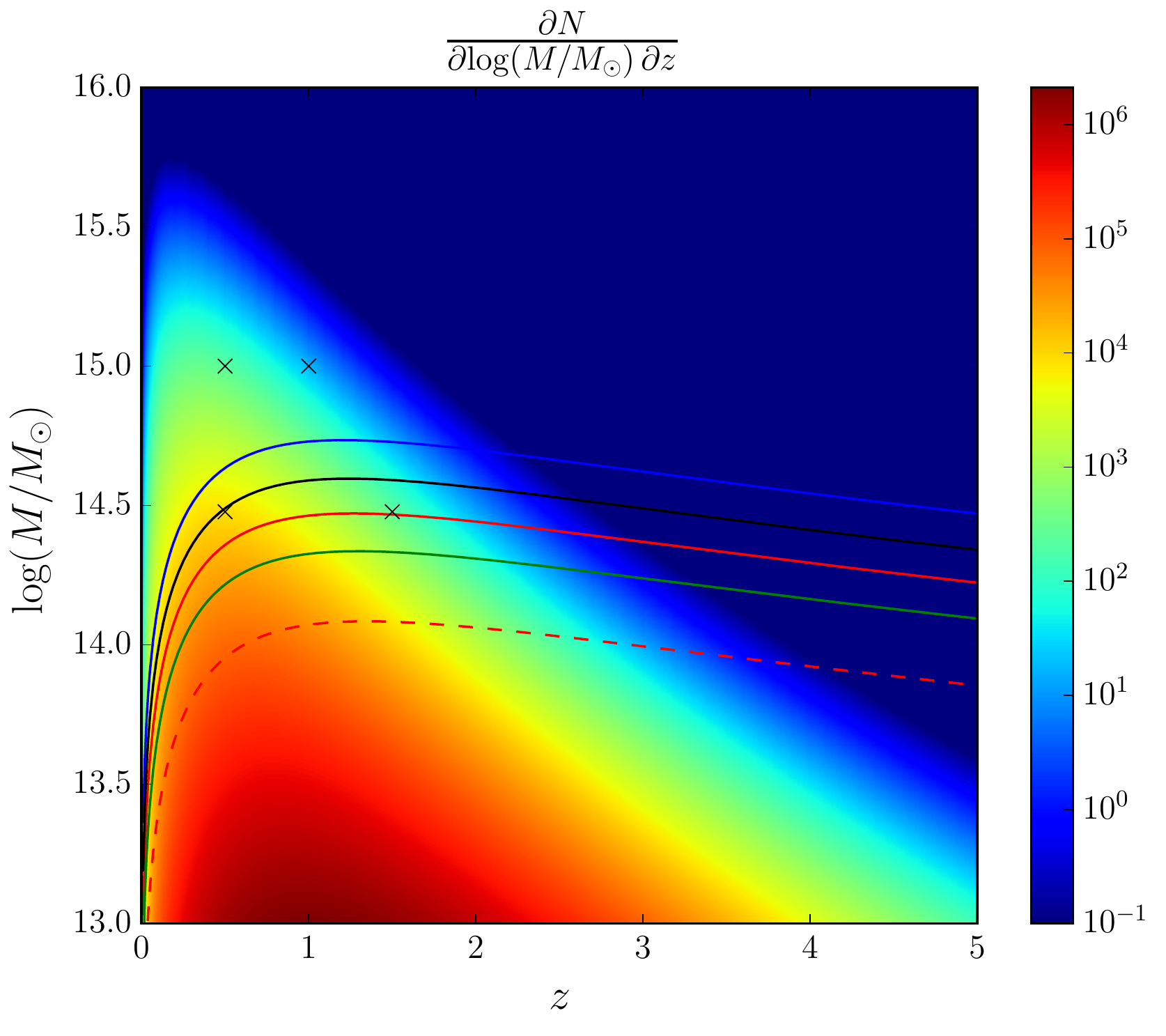}
\includegraphics[width=.5\textwidth]{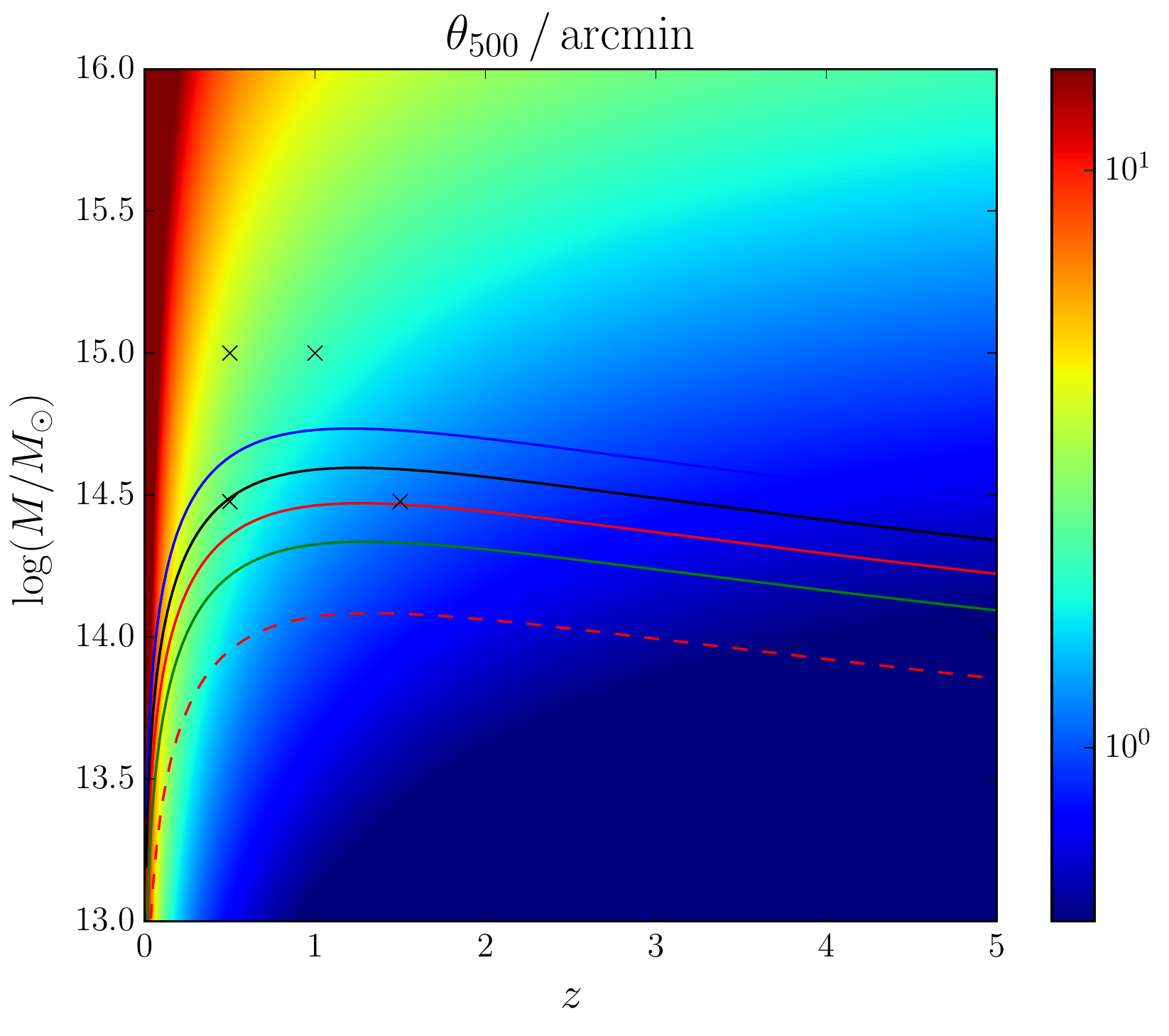}
\includegraphics[width=0.5\textwidth]{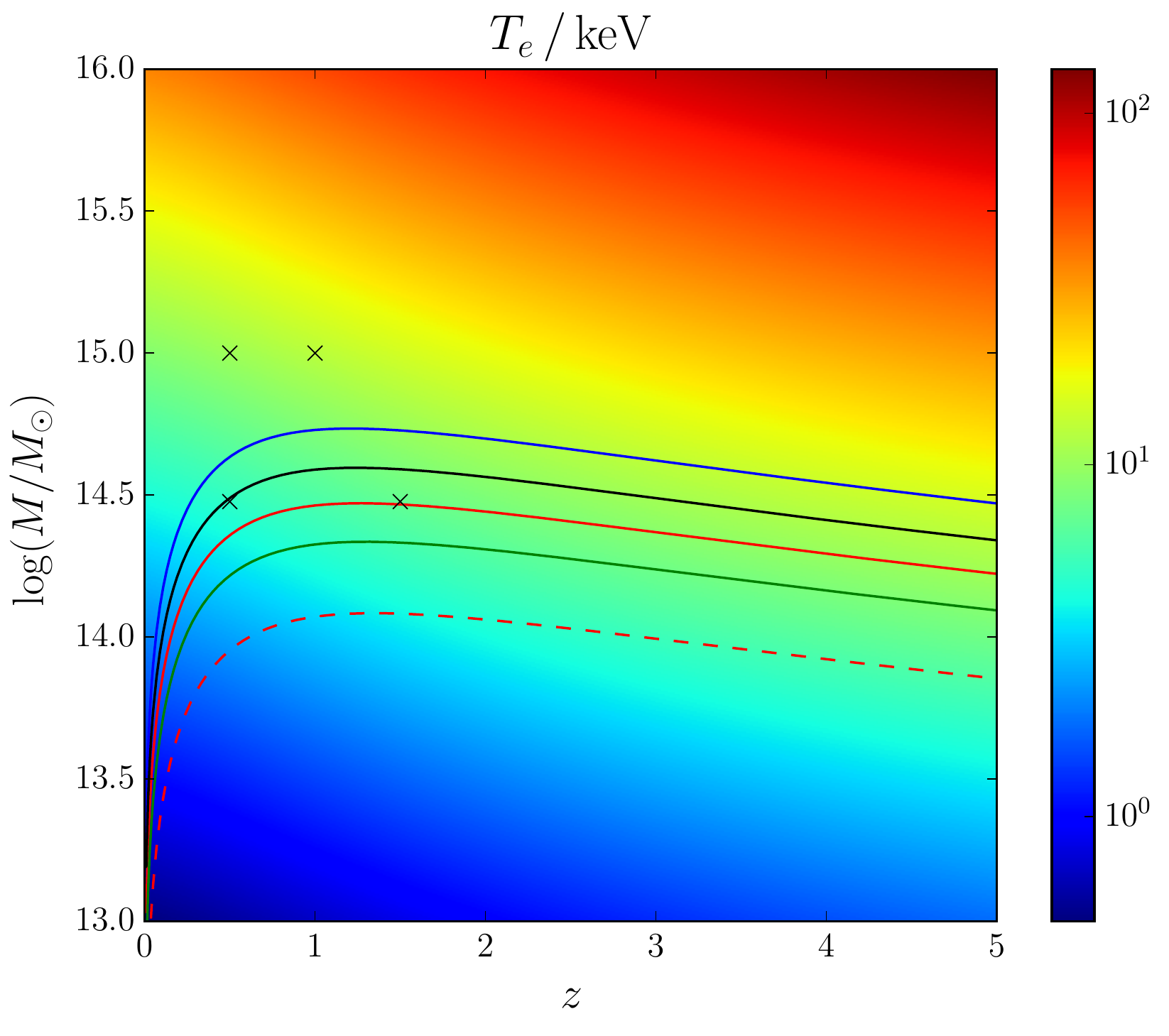}
\includegraphics[width=0.5\textwidth]{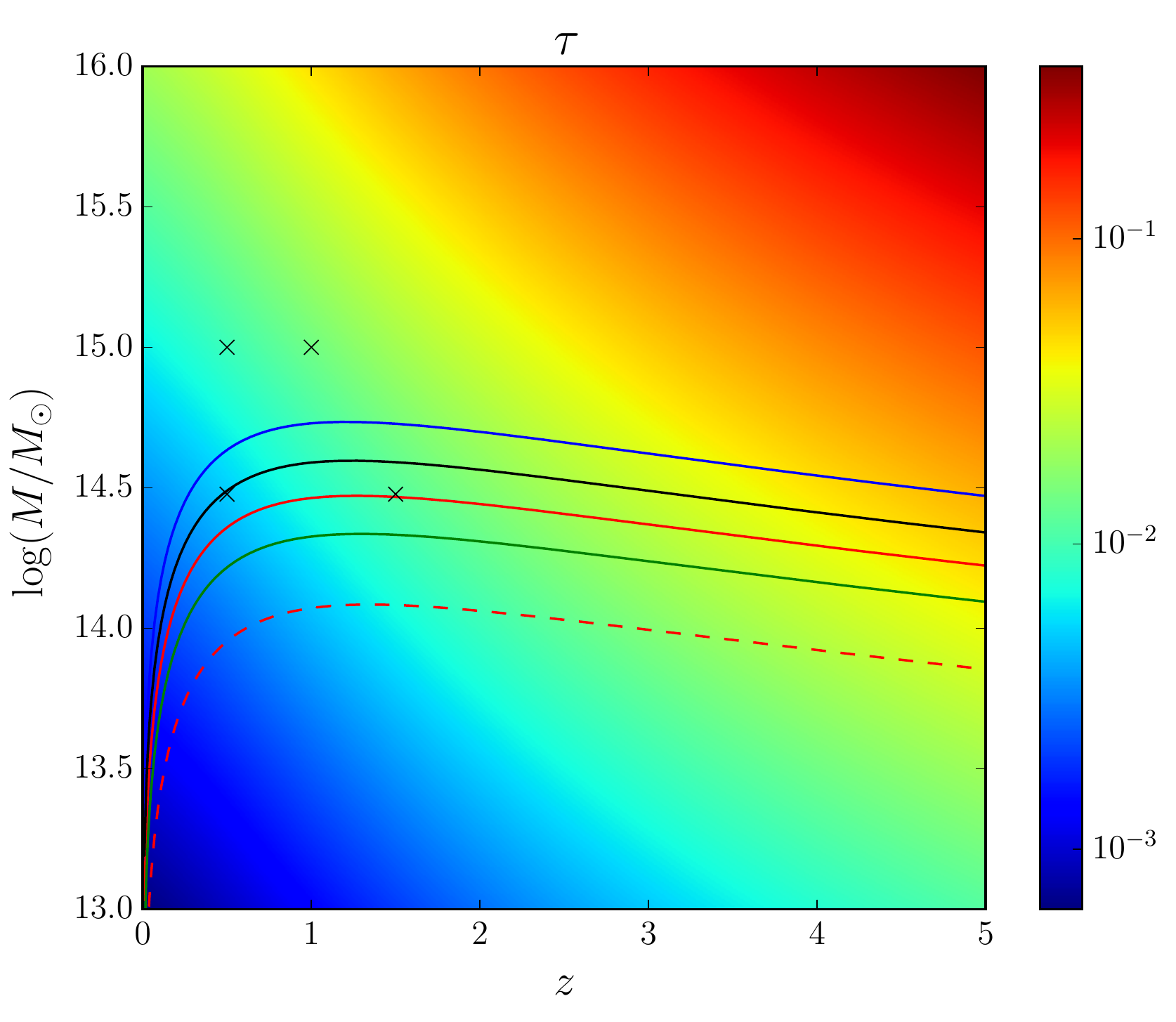}
\caption{Dependence of the number density (top-left), angular size $\tfoo$ (top-right), optical depth $\tau$ (bottom-right) and electron temperature $T_e$ (bottom-left) of clusters on their mass and redshift, calculated using the halo mass distribution from \cite{jenkins}, the $T$-$M$ scaling relation from \cite{2016MNRAS.463.3582M} given in equation~\ref{TM}, and the $Y$-$M$ scaling relation from \cite{Ade:2013lmv} given in equation~\ref{YM}. The number density plot assumes a survey area of $10^3$ deg$^2$, and a $10^4$ deg$^2$ survey will see $10\times$ more clusters than indicated by the colorbar. The minimum mass detection cutoffs are overlaid on all plots, with the lines progressing from top-to-bottom showing cutoffs for $10^4$~deg$^2$ at 4,000~hr (blue) and 16,000~hr (black), then $10^3$~deg$^2$ at 4,000~hr (red) and 16,000~hr (green), and then CCAT$_{opt}$ (red dashed, like in Figure \ref{distr}). The black crosses indicate the example clusters presented in Figures \ref{contour_op1} and \ref{contour_op4}. The uncertainties of $P_1,P_2,P_3$ depend only on $\tfoo$, and the transformation to $T_e,\tau,v$ introduces dependences on $T_e$ and $\tau$ (ref. \S\ref{fisher}).\label{mz_n_t500}}
\end{figure*}

For a given redshift, we find the minimum mass observable, by imposing a 5-$\sigma$ detection criterion, i.e. we consider as detected all the clusters for which the signal within an aperture of $\tfoo$ is at least 5 times the noise. We then use the Jenkins halo mass distribution \citep{jenkins} to find the number of clusters at each redshift with mass above the detection cutoff. The redshift and mass distribution of halos for a $10^3$~deg$^2$ field is shown in Figure \ref{mz_n_t500}. The figure also depicts the cluster size $\tfoo$, electron temperature $T_e$, and optical depth $\tau$ as functions of $M$ and $z$ (ref. equations \ref{TM} and \ref{YM}). The detection cutoffs for the four CCAT$_{base}$ strategies as well as CCAT$_{opt}$ are overlaid on the plots, and the cluster parameters used in Figures \ref{contour_op1} and \ref{contour_op4} are marked.

Since the only cluster parameter that changes the derivatives of the signal (equation \ref{sz_signal}) is $\tfoo$, the uncertainties in $P_1,P_2,P_3$ are functions solely of $\tfoo$. Larger clusters have signals that can be integrated over a greater area, so we would expect uncertainties to improve with $\tfoo$, which is indeed what we find. The transformation to $T_e,\tau,v$ introduces dependencies on $T_e$ and $\tau$, while only the velocity uncertainty $\sigma(v)$ depends on $v$ (ref. equation~\ref{cov_trans_matrix}). Equation~\ref{fom_trans} shows that the FoM in $T_e,\tau,v$ must be proportional to $T_e^2\tau^2f(\tfoo)$, where $f$ is a monotonically increasing function. One would expect all uncertainties to improve with each of $\tfoo,T_e,\tau$ since $T_e$ and $\tau$ determine the strength of the signal, and $\tfoo$ determines its spatial extent.

\begin{figure*}
\includegraphics[width=0.5\textwidth]{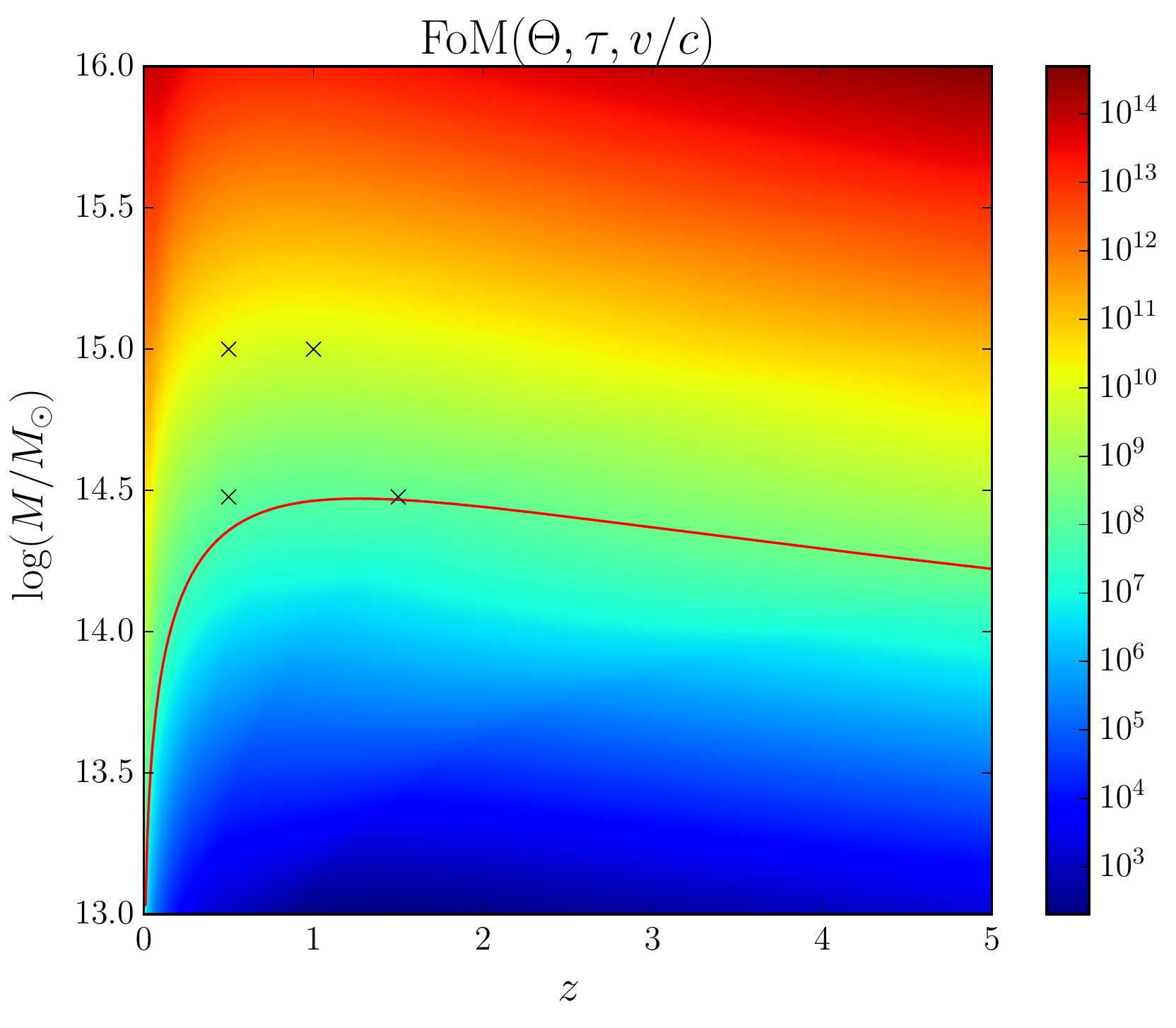}
\includegraphics[width=0.5\textwidth]{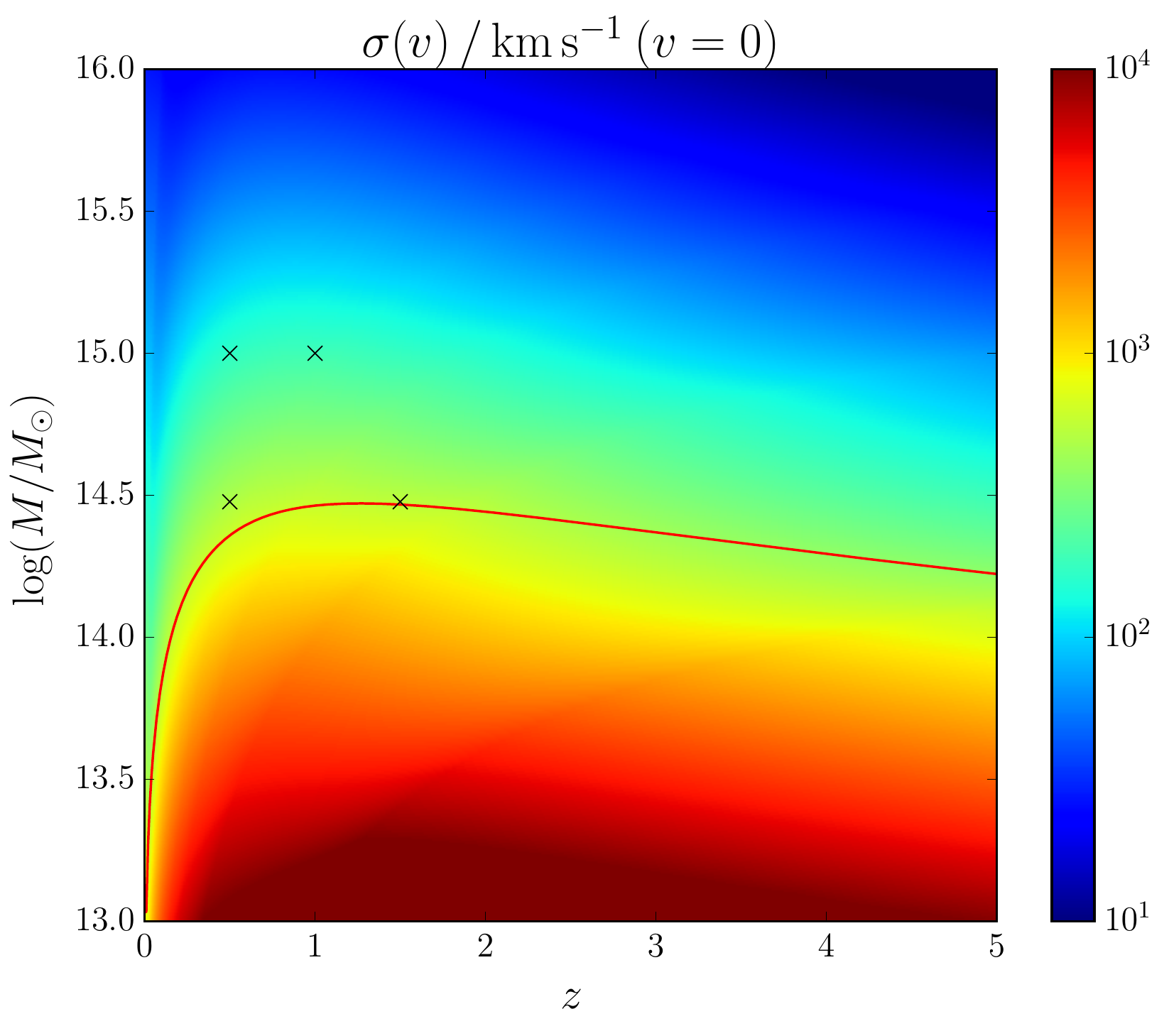}
\includegraphics[width=0.5\textwidth]{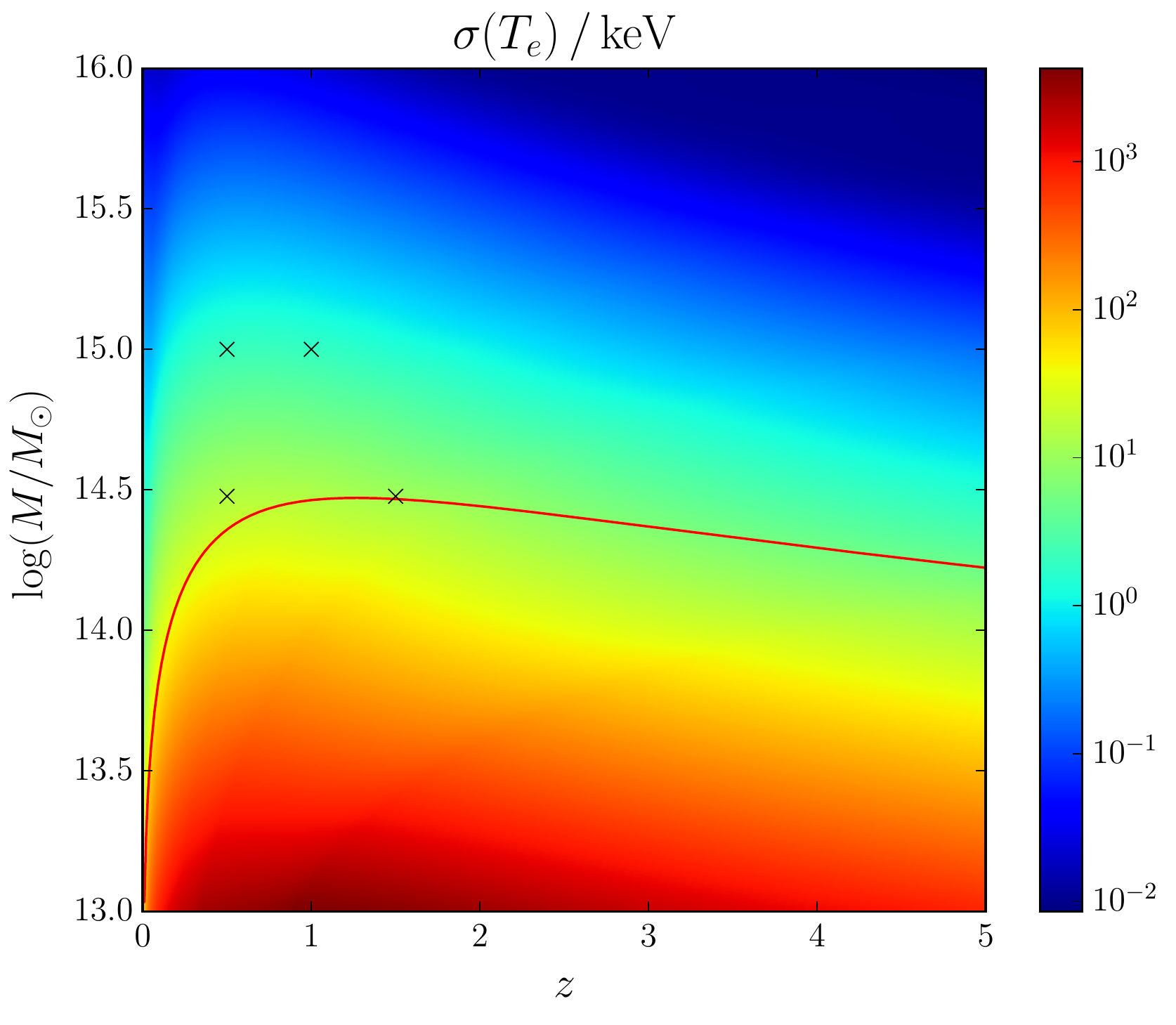}
\includegraphics[width=0.5\textwidth]{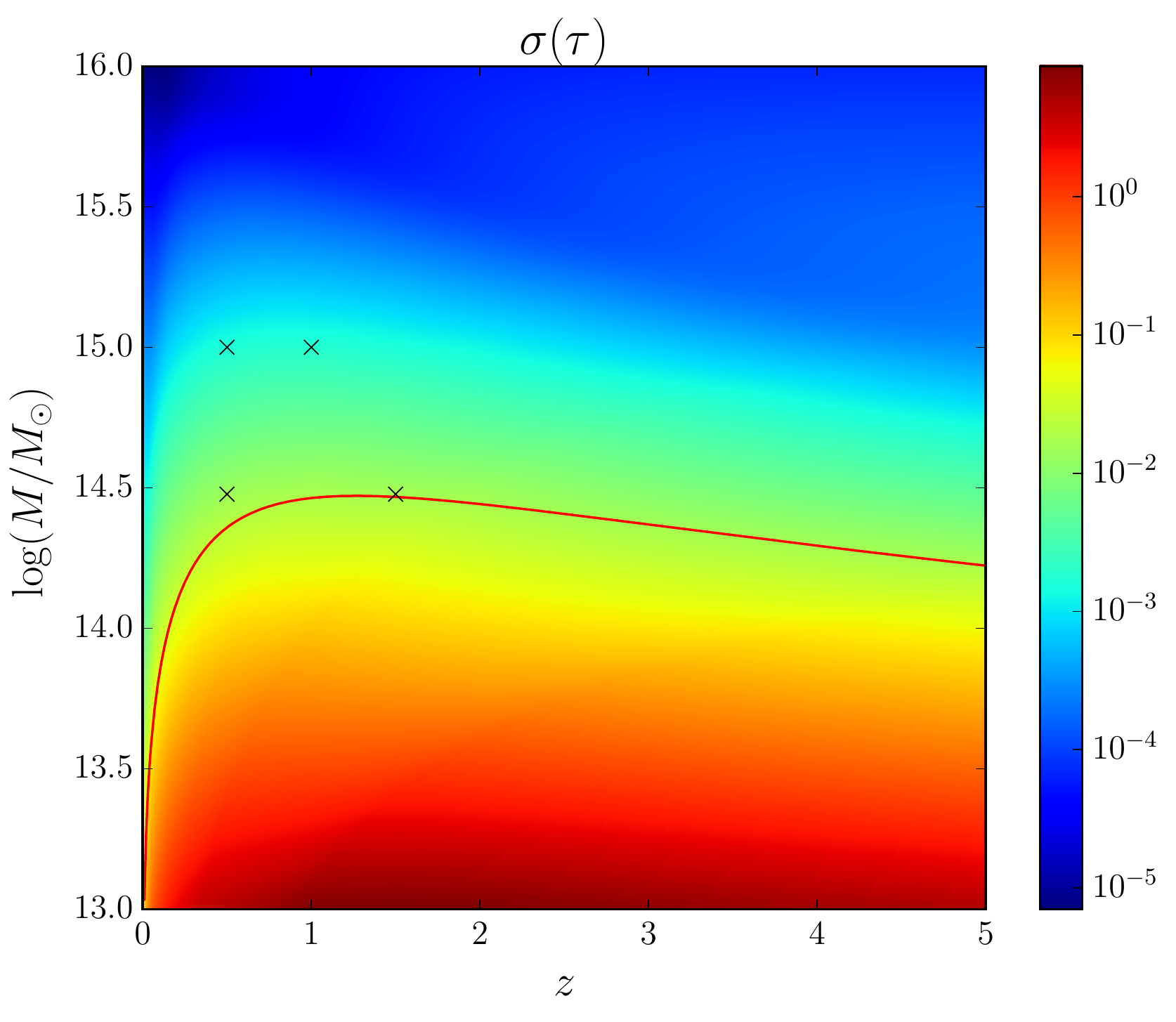}
\caption{Plots of uncertainties on cluster parameters as a function of their mass and redshift, for the baseline CCAT-p strategy (CCAT$_{base}$, Table~\ref{expt_param}). Clockwise from top-left, the plots show the figure of merit (ref. \S\ref{fom}), velocity uncertainty $\sigma(v)$ (for fiducial $v=0$), temperature uncertainty $\sigma(T_e)$, and optical depth uncertainty $\sigma(\tau)$. The red curve indicates the minimum mass detection cutoff, and the black crosses are the results in Figure~\ref{contour_op1}. The small discontinuities are numerical effects from $\tfoo$ sampling, which is why they seem to follow $\tfoo$ contours (ref. Figure~\ref{mz_n_t500}). The velocity uncertainty is the only quantity that depends on the cluster velocity as a result of equation~\ref{cov_trans_matrix}. The uncertainties improve strongly with mass and mildly with redshift due to the variations of $\tfoo$, $T_e$, and $\tau$ presented in Figure~\ref{mz_n_t500}. \label{mz_results}}
\end{figure*}

Figure \ref{mz_results} depicts the bounds on cluster parameters obtained by the baseline CCAT-p strategy described in Table~\ref{expt_param}. It shows the FoM and uncertainties in $T_e$, $\tau$, and (for clusters with 0 peculiar velocity) $v$, along with the detection cutoff. The uncertainties depend in the expected ways on the cluster parameters $\tfoo$, $T_e$, and $\tau$. As can be seen in Figure~\ref{mz_n_t500}, all cluster parameters increase with mass, so the uncertainties improve quickly with mass as well. $\tfoo$ decreases with redshift, first rapidly and then slowly, while the other parameters increase strongly. Therefore $\tfoo$ dominates at low $z$ and the other parameters dominate at high $z$, with the result that uncertainties quickly deteriorate with redshift before slowly improving beyond $z\sim 0.5$.

The similarity of uncertainty contours with the detection cutoff in Figure~\ref{mz_results} demonstrates that the behaviour of the uncertainties is well-approximated by the single-band detection criterion, which depends on only $y$ and $\tfoo$. Due to the steep improvement in uncertainties with mass, and steep deterioration with redshift for low $z$, the clusters with good measurements are biased toward higher masses and lower redshifts, which should be accounted for in cosmological analysis.

\begin{figure}
\begin{center}
\includegraphics[width=\columnwidth]{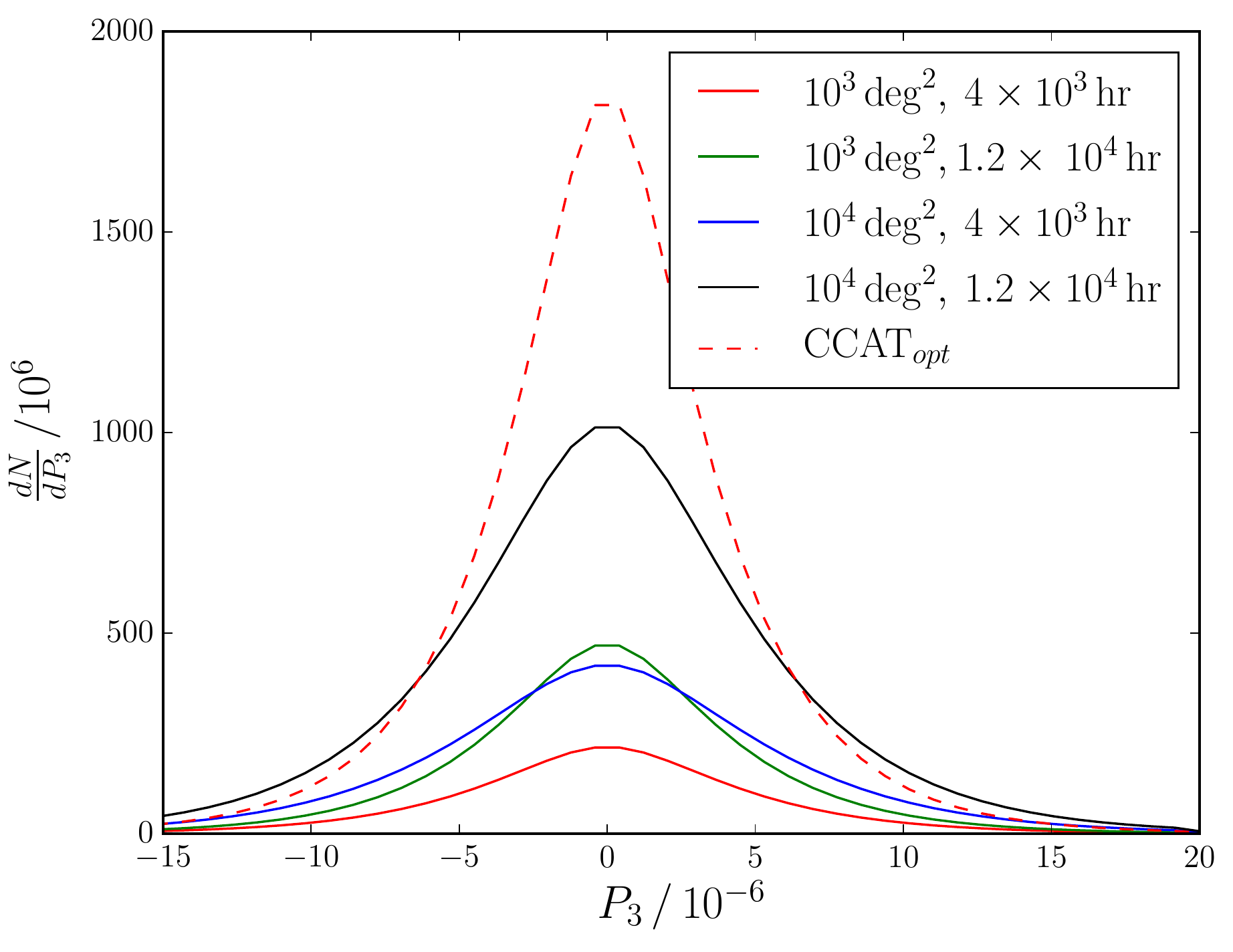}
\end{center}
\caption{The distribution of the kSZ amplitude $P_3$ (ref. equation \ref{p1p2p3}) of detected clusters for the 5 CCAT-p survey strategies. The standard deviation of each distribution is between $4.6 \times 10^{-6}$ and $6.3\times 10^{-6}$. This is important to consider when comparing to uncertainties in Figures \ref{survey_strats} and \ref{n_sigma}. 
The plot also shows that CCAT$_{opt}$ will detect far more clusters than the scaled CCAT$_{base}$ strategies, which is not surprising due to the expected sensitivity improvement. 
\label{distr}}
\end{figure}

We assume that the peculiar velocities are independent from $M,z$ and follow a Gaussian distribution with standard deviation $\sigma = 351\rm \, km\,s^{-1}$, based on \citet{Sheth:2000ii}. In Figure~\ref{distr} we present the expected distribution of the kSZ signal amplitude for the thousands of clusters detected by the five strategies. 
Since the sensitivity scales like $T^{\rm sens } \propto \sqrt{A/t}$, the detection threshold is increased for wider surveys, as shown in Figure \ref{mz_n_t500}. However in terms of the total number of clusters detected, this effect is dominated by that of seeing more of the sky with wider surveys, with the result that wider surveys detect more clusters.

\begin{figure}
\includegraphics[width=\columnwidth]{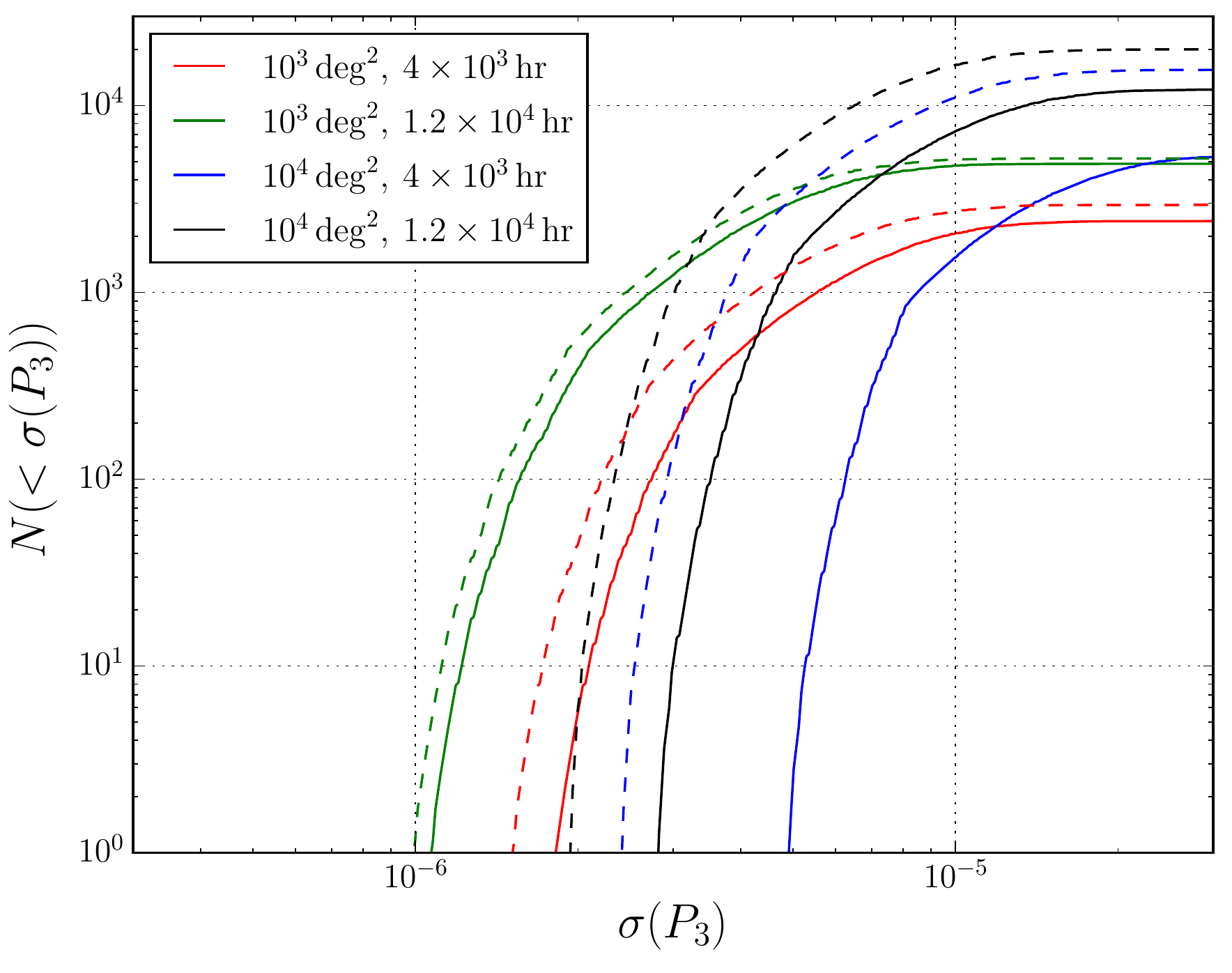}
\caption{Cumulative distribution of observed numbers of clusters with kSZ uncertainties for the four survey strategies based on the baseline CCAT-p instrument configuration. The $y$-axis represents the number of clusters that can be detected to within the uncertainty given by the $x$-axis. Solid lines represent results from CCAT-p, while dashed lines represent results from coadding maps from CCAT-p and Advanced ACTPol \citep{Henderson:2015nzj}. These uncertainties may be compared with the expected distribution of $P_3$ signals, which is Figure \ref{distr}. Small peaks and discontinuities are numerical effects from $z$ sampling.\label{survey_strats}}
\end{figure}

Figure \ref{survey_strats} displays the distribution of kSZ uncertainties of the detected clusters for the four CCAT-p survey strategies based on the CCAT$_{base}$ instrument configuration. The lines plateau to a constant value as a result of our 5-$\sigma$ tSZ detection criterion; in practice observing previously known clusters (e.g. from optical catalogs) would give us low-S/N measurements of several more clusters. The small peaks that are visible in the plots are numerical effects caused by $z$ sampling at low redshift; the true results should be smooth. 

As expected, we find that (for fixed survey length) deeper surveys see fewer clusters but to higher significance, while wider surveys detect more clusters with higher uncertainties. Thus the choice of survey depth represents a trade-off between quality and quantity, though this effect may be diluted by the slightly lower signal amplitudes that are detected, on average, by deeper surveys (ref. Figure~\ref{distr}). In addition, deeper surveys explore more of the $M,z$ parameter space than broader ones do (ref. Figure~\ref{mz_n_t500}).

The dashed lines in Figure~\ref{survey_strats} show the results from coadding the CCAT-p maps with overlapping measurements from the Advanced ACTPol (AdvACT) project \citep{Henderson:2015nzj}. Most of the low-foreground sky will be observed with AdvACT with similar resolution between 90 GHz and 230 GHz; therefore  we can simply coadd the AdvACT sensitivity estimates with those of CCAT-p. The improvement from coadding the AdvACT data is strongest for the wider, shallower surveys due to the larger area and higher level of map noise of AdvACT. For the CCAT-p baseline strategies, adding AdvACT results in a 20--70\% increase in the number of clusters detected to a given uncertainty. 

\begin{figure*}
\includegraphics[width=.5\textwidth]{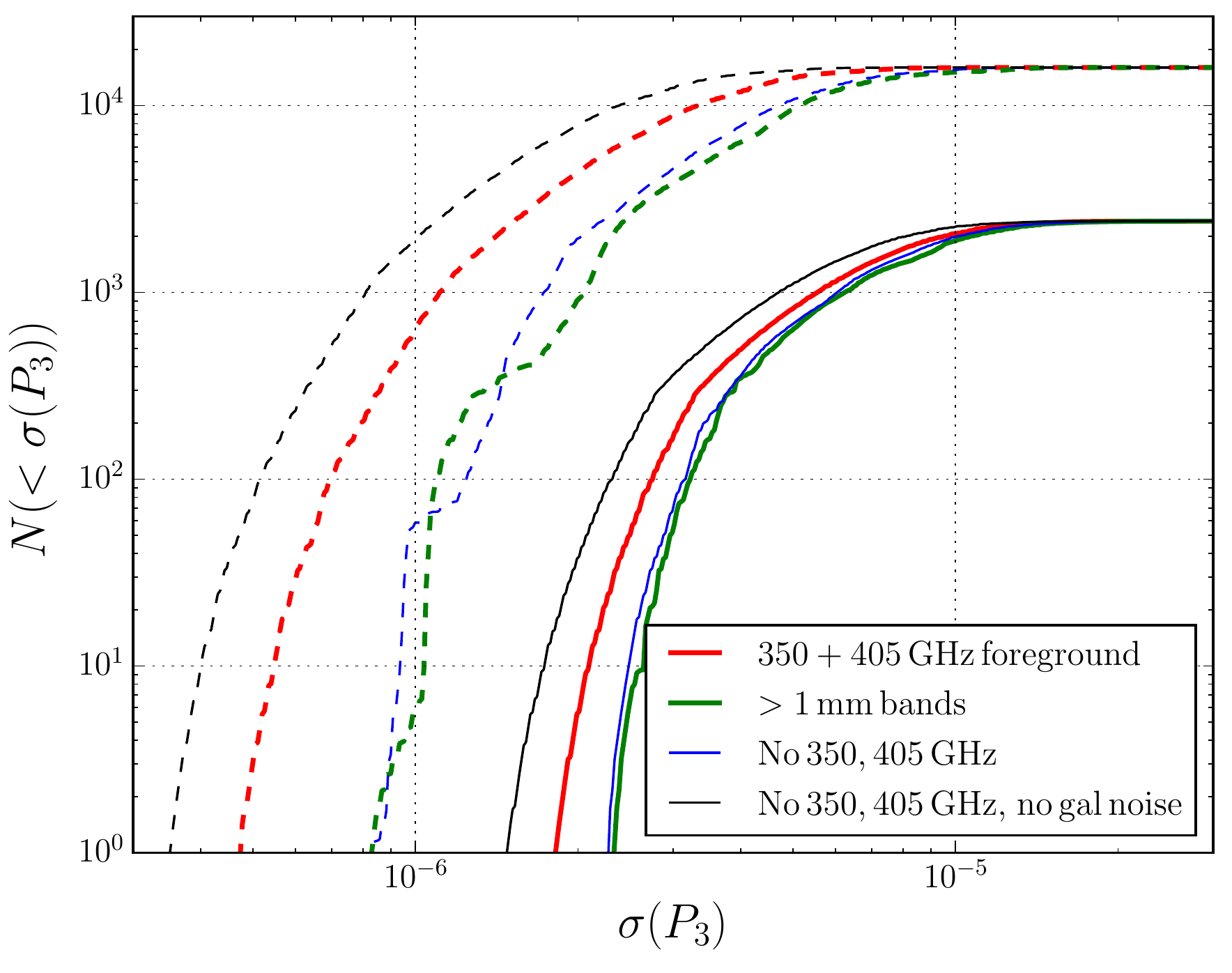}
\includegraphics[width=.5\textwidth]{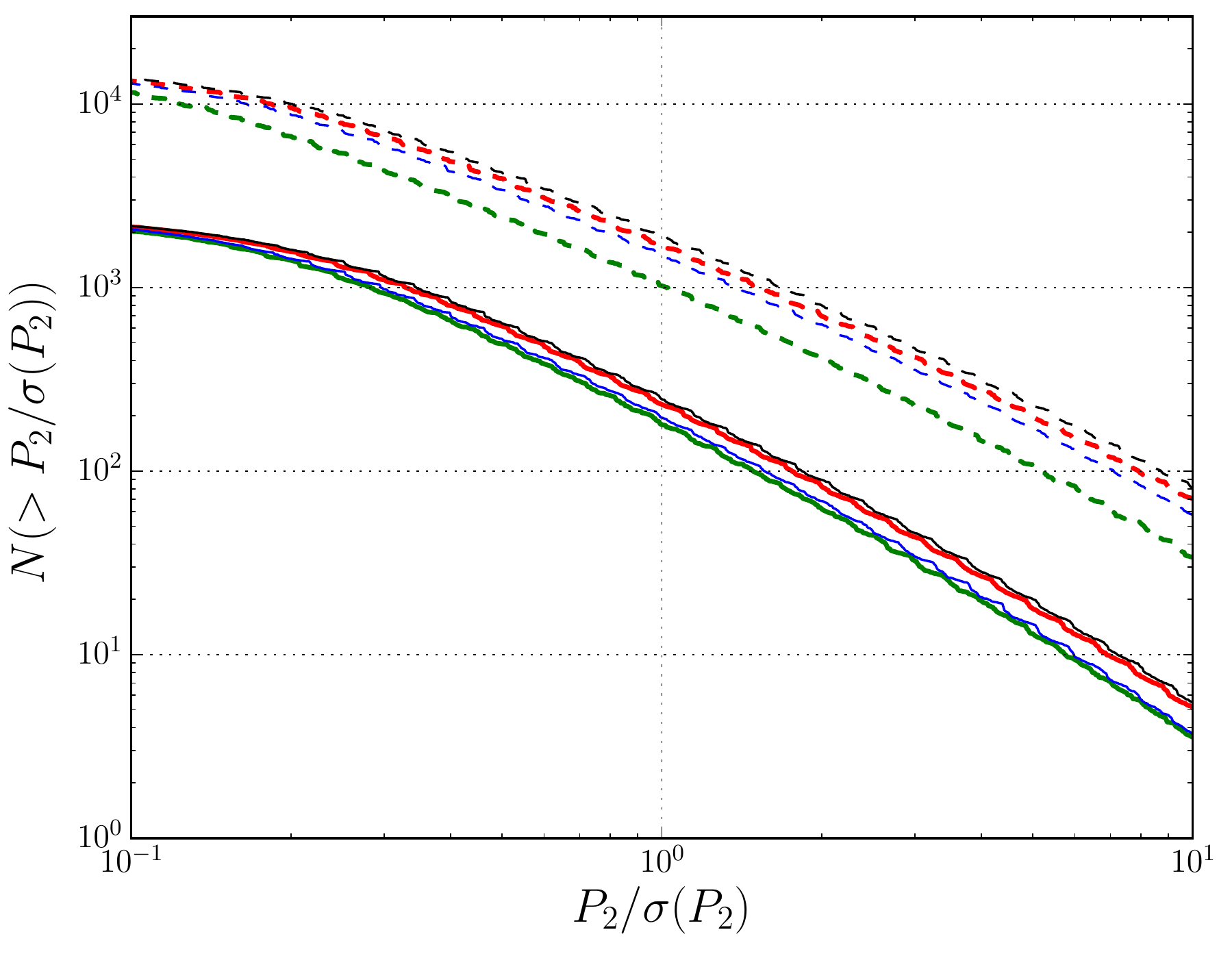}
\includegraphics[width=.5\textwidth]{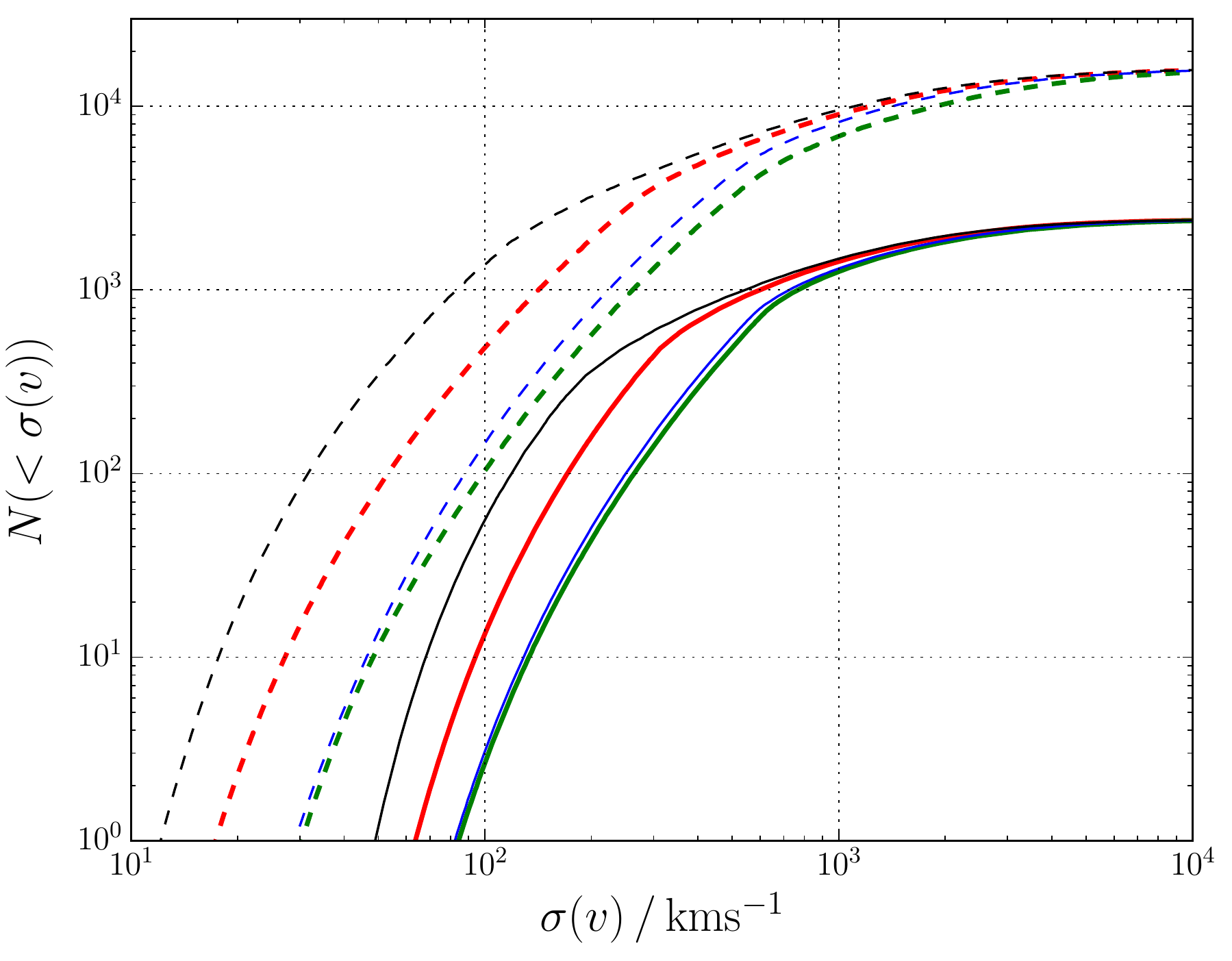} 
\includegraphics[width=.5\textwidth]{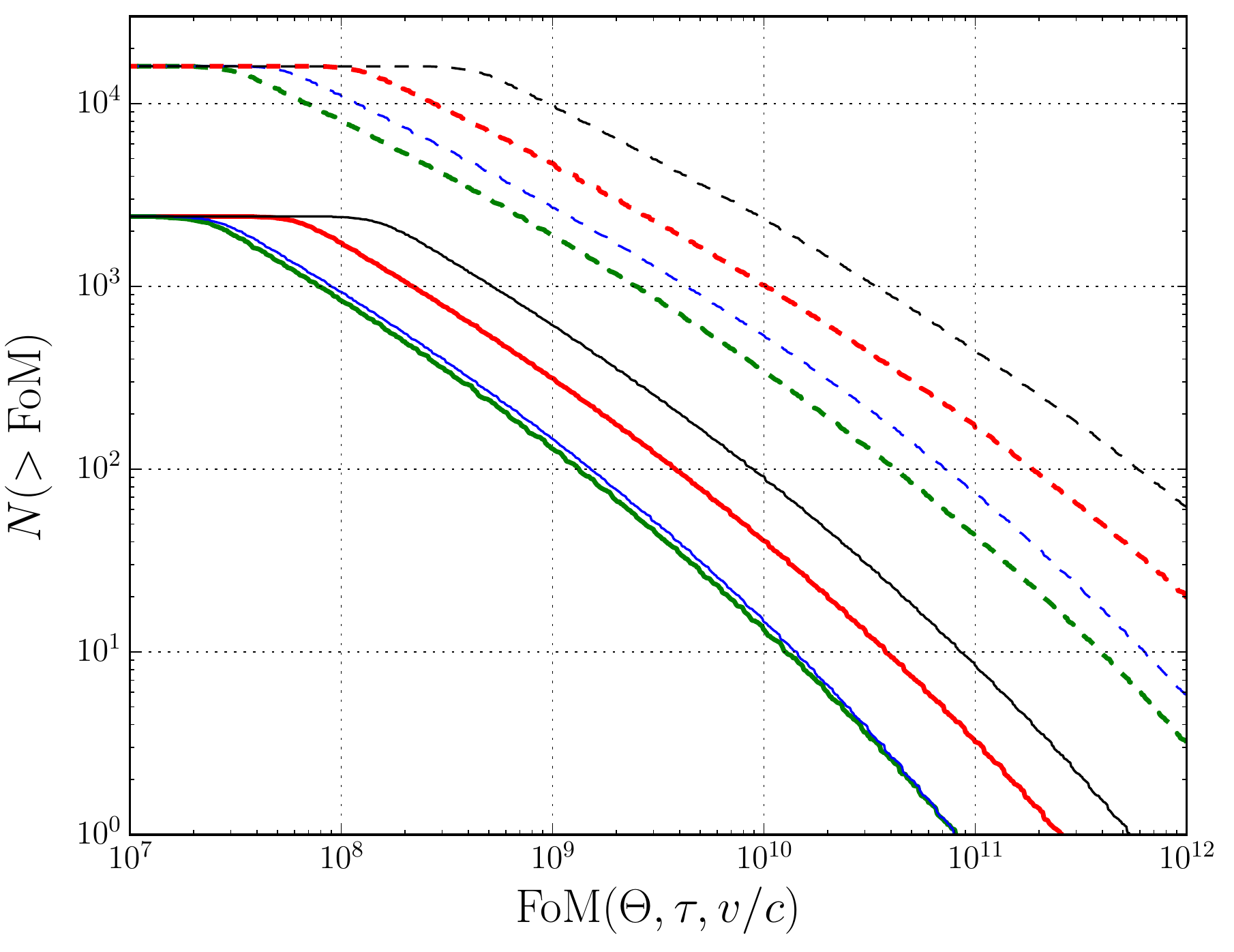}
\caption{Cumulative distribution of observed numbers of clusters with uncertainties for the CCAT-p strategies described in Table \ref{expt_param}. The $y$-axis represents the number of clusters that can be detected to the uncertainty, signal-to-noise, or FoM given by the $x$-axis or better. (Note that better measurements are represented by lower $\sigma$ but higher S/N and FoM.) Solid lines represent results from CCAT$_{base}$, while dashed lines represent results from CCAT$_{opt}$. Clockwise from top-left, the uncertainties plotted are: kSZ signal amplitude uncertainty, rSZ signal-to-noise ratio, FoM, velocity uncertainty. 
The curves representing the standard CCAT configurations (red) and that without submillimeter bands (green) are emphasised by thicker lines. Comparing the two shows that submillimeter bands increase the number of clusters detected to a given uncertainty by a factor of $\sim3$. Analysis with no 350 and 405 GHz foreground removal channels (blue lines) can also be compared to similar analysis with no DSFG noise (black lines), which shows the effectiveness of the foreground subtraction (red lines). With increased DSFG noise (see \S\ref{sec:dsfgs}) the gap grows between the default configuration lines (red) and configurations without submillimeter information (blue and green). Thus, the importance of submillimeter bands and foreground subtraction may be greater than this figure suggests. Comparing the dotted and dashed curves demonstrates that CCAT$_{opt}$ will see up to 10$\times$ more clusters to any given uncertainty than CCAT-p, and could measure the peculiar velocities of hundreds of clusters to within 100 km/s. Small discontinuities, peaks, and troughs are numerical effects.\label{n_sigma}}
\end{figure*}

Next, we analyze the impact of submillimeter bands and foreground subtraction. Figure \ref{n_sigma} depicts distributions of uncertainties for the CCAT$_{base}$ and CCAT$_{opt}$ surveys. It also shows the distributions when different combinations of the higher-frequency bands are ignored. Comparing these curves demonstrates the importance of submillimeter bands, which can increase the number of clusters detected to a given uncertainty by factors $> 3$ in some regimes. The cases when both foreground bands (350 and 405~GHz) are ignored, with and without DSFG noise, are included in the figure. These present an upper and lower bound to the foreground-subtracted default configuration, and allow us to assess the effectiveness of foreground subtraction. We see that with the high frequency bands we are able to recover velocities with small $\sigma(v)$ for up to 70\% of the clusters contaminated by significant DSFG noise.

Figure \ref{n_sigma} also presents the same information for CCAT$_{opt}$, the 16,000 hr survey with an upgraded instrument. This shows that CCAT$_{opt}$ will measure kSZ signals for over 10$\times$ more clusters to any given uncertainty than the CCAT$_{base}$ first light instrument configuration, and that high-frequency bands (particularly the one at 862~GHz) contribute significantly to this improvement.

\subsection{Optimizing experiments}\label{sec:pair_det}

In the previous section we examined changes in survey strategy. In this section we vary experimental parameters to optimize the SZ FoM.
As described in \S\ref{experiments}, the CCAT$_{base}$ configuration evolved from the SWCam instrument design \citep{stacey/etal:2014}, which included seven independent optics tubes installed in one cryogenic instrument. The seven optics tubes were assumed to illuminate six dual-frequency arrays (two at 95/150 GHz, two at 226/273 GHz, two at 350/405 GHz) and one single-frequency array (862 GHz). We are also exploring instrument concepts with as many as 50 independent optics tubes in one cryogenic instrument \citep{niemack:2016}. For the instrument optimization process, we treat each of the seven frequencies independently, which is analogous to having 13 single frequency optics tubes (two each at 90, 150, 226, 273, 350, 405 GHz and one at 862 GHz, Table \ref{expt_param}) and optimizing between them.
The CCAT$_{opt}$ configuration was optimized by allowing up to 50 single frequency optics tubes; however, as we describe later, the CCAT$_{opt}$ configuration could be deployed in as few as 28 optics tubes by using dual-frequency detector arrays at frequencies $<$ 500 GHz.

\begin{figure}
\includegraphics[width=1\columnwidth]{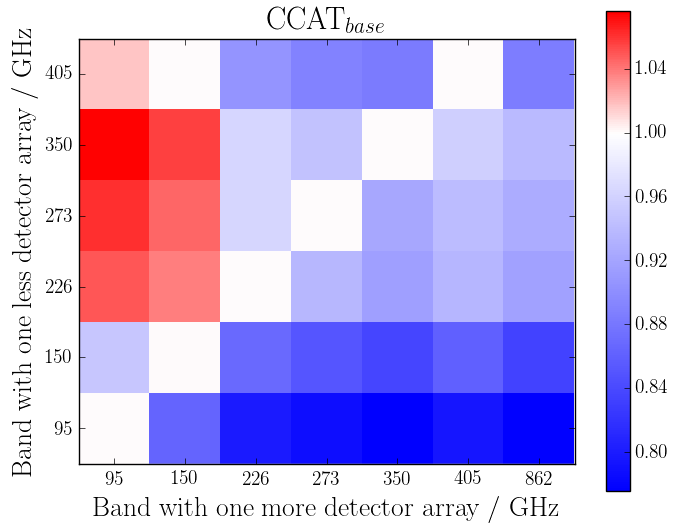}
\includegraphics[width=1\columnwidth]{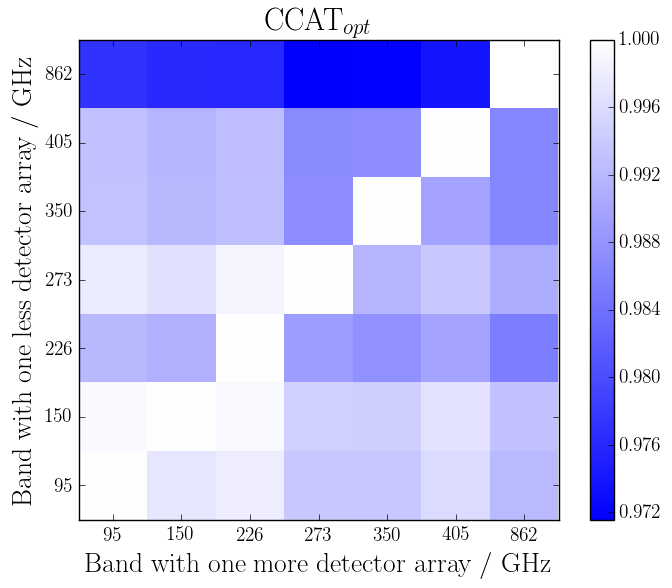}
\caption{\textit{Top:} The FoM matrix, normalised to the diagonal, for moving one detector array between each pair of bands (ref. \S\ref{sec:pair_det}) for the baseline CCAT$_{base}$ configuration. The row corresponding to removing an array from the 862~GHz band is omitted since it has only one such array, and we do not allow this to go to zero in the optimization. The baseline CCAT-p configuration would benefit from having more arrays in the lower frequency bands. \textit{Bottom:} Same as top panel, for the optimized CCAT$_{opt}$ configuration with a total of roughly 50 single-frequency detector arrays or roughly 28 dual-frequency arrays. The configurations are summarized in Table~\ref{bands_opt}. CCAT$_{opt}$, as its name suggests, has already been optimised. \label{bands_matrix}}
\end{figure}

To perform the optimization, we compute the matrix of FoMs after transferring one array between each pair of bands.
We do not expect this to depend heavily on the cluster parameters, and have verified that multiple cluster parameters converge to the same configuration. Thus we only present results for the fiducial cluster parameters presented in Table \ref{fiducial_cluster}. Examples of such matrices are given in Figure \ref{bands_matrix}. If, for a pair of bands, moving an array either way results in a loss of FoM, then the array assignment between these two bands is locally optimal. By iteratively changing the array assignment based on the FoM matrix, and recomputing the matrix, we may find such an optimal allocation of detector arrays. 

We have done this for a baseline CCAT$_{base}$ configuration and for a next generation CCAT$_{opt}$ upgrade with up to 50 single frequency optics tubes, instead of 13. We keep the total number of bands fixed, i.e. we impose that each band must have at least one detector array, although we find that all bands are allocated more than one optics tube in the CCAT$_{opt}$ configuration. The results are presented in Figure \ref{bands_matrix} and Table \ref{bands_opt}. The most important bands are 95 and 150 GHz: detector arrays are allocated to 31 of the 50 optics tubes at these frequencies.
The submillimetre bands are allocated 11 arrays, and all bands find an equilibrium with more than one array, showing that they all provide value, and emphasising the effectiveness of having multiple frequencies.

\begin{deluxetable*}{c|ccccccc|c}
\tablecaption{The considered CCAT-p configurations before and after optimizing the assignment of detector arrays to frequency bands for multiple sets of fiducial cluster parameters. The rule of thumb is motivated by the fact that the band ratios remain roughly constant. The FoMs presented are for the fiducial cluster described in Table \ref{fiducial_cluster}. We have verified that various cluster parameters converge to the same optimum detector array allocation.
\label{bands_opt}}
\tablehead{Configuration & \multicolumn{7}{|c|}{Number of detector arrays (sensitivity$/\mu$K arcmin)} & FoM($\Theta,\tau,v/c$)\\ & 95 GHz & 150 GHz & 226 GHz & 273 GHz & 350 GHz & 405 GHz & 862 GHz & $/ 10^9$}
\startdata
CCAT$_{base}$ baseline (ref. Table \ref{expt_param}) & 2 (4.9) & 2 (6.4) & 2 (4.9) & 2 (6.2) & 2 (25) & 2 (72) & 1 ($6.6\times 10^4$) & 2.31 \\
CCAT-p baseline optimized & 4 (3.5) & 4 (4.5) & 1 (6.9) & 1 (8.8) & 1 (36) & 1 (100) & 1 ($6.6\times 10^4$) & 2.73 \\
Optimized with half arrays & 4 (3.5) & 3.5 (4.8) & 1.5 (5.7) & 1 (8.8) & 1 (36) & 1.5 (83) & 0.5 ($9.4\times 10^4$) & 2.81 \\
\rule[-1.5ex]{0pt}{0pt}CCAT$_{opt}$ (ref. Table \ref{expt_param}) & 16 (0.9) & 15 (1.2) & 3 (2.0) & 5 (2.0) & 4 (8.9) & 5 (23) & 2 ($2.3 \times 10^4$) & 107\\
\hline
\rule{0pt}{2.75ex}Throughput ratio rule of thumb & 4 & 4 & 1 & 1 & 1 & 1 & 0.5
\enddata
\end{deluxetable*}

\section{Conclusion}\label{conclusion}

We present Fisher matrix forecasts of the properties and uncertainties of galaxy clusters that CCAT-p will be able to detect. 
Larger area surveys detect more clusters, but with larger uncertainties on the cluster properties than deeper surveys. We find that submillimeter bands play an important role in extracting SZ signals and eliminating noise from dusty star-forming galaxies. 

We also use Fisher matrices to optimize the frequency balance of detector arrays for a CCAT-p first light instrument concept and a CMB-S4 scale instrument concept. We forecast the results of a futuristic (optimized) upgrade to CCAT-p called CCAT$_{opt}$, and find that with $\sim 4 \times$ more detectors, we could see up to $\sim 10 \times$ more clusters to any given uncertainty, measuring the peculiar velocities of thousands of clusters. Our results highlight the importance of submillimeter bands and foreground subtraction for cluster measurements.

We find that optimized ratios of the number of detector arrays at each frequency do not change significantly between the CCAT-p first light and CMB-S4 scale instrument concepts. If multiple high throughput six-meter aperture telescopes are built for CMB-S4, this suggests that extraction of the cluster parameters $T_e$, $\tau$, and $v$ may be optimized by continuing to pursue a similar frequency balance to the rule of thumb in Table \ref{bands_opt}. Of course, there are many other drivers of the distribution of frequencies to consider for CMB-S4; however, this analysis suggests that significant improvements in the CMB-S4 galaxy cluster science may be achieved by adding modest submillimeter capabilities on a telescope like CCAT-p.  

Individual cluster measurements at the level of precision achievable by CCAT-p will provide insight into interesting areas of cluster astrophysics, such as temperature profiles, turbulent flows, and AGN feedback. Velocity measurements for a range of cluster masses and redshifts will provide constraints on the cosmic velocity field, and yield insight into the growth of large-scale structure and cosmology. For example, \cite{2007ApJ...659L..83B} show that measuring the peculiar velocities of all clusters with $M> 10^{14} M_\odot$ to 100 km s$^{-1}$ for a 5000 deg$^2$ survey would constrain each of the Hubble constant, the primordial power spectrum index, the normalization of the matter power spectrum, and the dark energy equation of state to better than 10\%, independent of other cosmological probes.

Future forecasts could build on these by including noise terms such as radio noise, dusty galactic emissions from within the cluster itself, and dust from our own Galaxy. These terms are only expected to increase the relative importance of broad frequency coverage. Another important direction for future analyses is extending the distribution of cluster parameters presented here to cosmological and astrophysical parameter constraints. Significant recent progress has been made along these lines building on the CMB-S4 science book; however, the majority of recent calculations do not consider submillimeter wavelengths like those that will be measured with CCAT-p. In other words, this work highlights one of the ways in which CCAT-p offers unique new galaxy cluster measurement capabilities as a potential platform for future CMB-S4 measurements.

\acknowledgments

We are grateful to members of the CCAT-prime collaboration as well as the Simons Observatory and CMB-S4 Collaborations for useful discussions. Gordon Stacey, Thomas Nikola, and Steve Parshley provided help with the instrument concepts and sensitivity estimates. We thank Colin Hill, Kaustuv Basu, Nick Battaglia, Jens Erler, Eve Vavagiakis, Douglas Scott, Rachel Bean, Frank Bertoldi, Joseph Mohr, Eiichiro Komatsu, Ted Macioce, and the referee for useful discussions and/or comments on the paper. The authors acknowledge support from the US National Science Foundation awards AST-1454881 and AST-1517049. 

\appendix

\section{Comparison with measurements and systematics}\label{sec:comparison}

To test the approach adopted in this paper, we compare the results of our forecasts with existing constraints on cluster parameters from recent measurements. Specifically, \citet{lindner} analyze the SZ properties of 11 galaxy clusters using  data from the Large APEX Bolometer Camera (LABOCA) on the Atacama Pathfinder Experiment (APEX), and data from the Atacama Cosmology Telescope (ACT). They use data from the Herschel Space Observatory's Spectral and Photometric Imaging Receiver (SPIRE) to subtract the dusty emission foreground, as described in \S\ref{foreground}. The relevant parameters of these experiments are described in Table \ref{expt_param2}. They also use the 2.1 GHz Australia Telescope Compact Array to remove radio sources, but we ignore that as we neglect radio sources in the analysis.

\citet{lindner} obtain prior X-ray measurements of $T_e$ for all but 2 clusters. Hence we exclude those 2 clusters from our analysis, leaving a total of 9 clusters. They obtain values of $\tfoo$ using the cluster radii and redshifts, then filter and clean the data, deconvolving them into continuous, roughly circular maps of radius $\sim 5'$.  They then measure the integrated SZ flux within an aperture $\theta'$, to find the integrated comptonization parameter $Y\equiv \int y\, d\Omega$ within this aperture, as well as the bulk peculiar velocities $v$ of the galaxy clusters. \citet{lindner} estimate that only a fraction (between 0.1--0.54) of the arcminute-scale SZ signal is reconstructed in their LABOCA data analysis, which is expected to significantly increase the uncertainties in their results. They report the measured parameters and the 1-$\sigma$ uncertainties, which we summarize in Table \ref{lindner_data}.

For each cluster, we set the fiducial value for the cluster parameters exactly at the measured values reported by \cite{lindner}, and forecast the uncertainties for the same map noise levels, which are summarized in Table \ref{expt_param2}. To get $y$, and thus $\tau$, from the reported integrated $Y'$ within a $\theta'$ radius we use the cluster model from \S\ref{SZ} and equation \ref{int_y_eq}.

Since $T_e$ is treated as a prior rather than a measurement, we compare only the forecast uncertainties in velocity $\sigma_{\rm f}(v)$ and optical depth $\sigma_{\rm f}(\tau)$ to the ones reported by \citet{lindner}. These uncertainties are presented in Table \ref{lindner_data}. We find the measured and forecast uncertainties to be correlated, with a Pearson correlation coefficient of 0.74 for $v$ and 0.88 for $\tau$, but the forecast uncertainties are systematically smaller than the measurement uncertainties by a factor of $\sim 2$. The dominant source of this discrepancy is believed to be the limited reconstruction of the arcminute-scale SZ signal in the LABOCA maps described above. \citet{lindner} also emphasize the importance of making multiple simultaneous measurements at millimeter and submillimeter wavelengths to extract these signals, which we plan to do with CCAT-prime.

\begin{deluxetable}{cccc}
\tablecaption{Experimental parameters from \cite{lindner}. Note that each cluster in 
Table \ref{lindner_data} has a different sensitivity, so here we present averages.
\label{expt_param2}}
\tablehead{\colhead{Name} & \colhead{$\nu$} & \colhead{$\xi$} & \colhead{$T^{\rm sens}(\rm avg)$} \\
 & GHz & arcmin & $\mu$K-arcmin}
\startdata
ACT & 148 & 1.4 & 53\\
     & 218 & 1.1 & 55\\
     LABOCA & 345 & 0.47 & 155
\enddata
\end{deluxetable}

\begin{deluxetable*}{c|ccc|ccccc|cc|cc}
\tablecaption{Cluster-wise data from the comparison of the forecasts to real results obtained by \citet{lindner}. From left to right, the columns are: cluster name in reference to the ACT catalogue, map sensitivities for each of the three bands, cluster angular radius from previous results, prior measurement of ICM temperature, error in prior temperature measurement, observed optical depth reported by Lindner, observed cluster velocity, error in observed optical depth, error in observed velocity, forecasted uncertainty in optical depth, forecasted uncertainty in velocity.
\label{lindner_data}}
\tablehead{\colhead{Name} &  \multicolumn{3}{c}{$T^{\rm sens}/\mu$K arcmin} & \colhead{$\tfoo$} & \colhead{$T_e$} & \colhead{$\sigma_{\rm p}(T_e)$} & \colhead{$\tau$} & \colhead{$v$} & \colhead{$\sigma(\tau)$} & \colhead{$\sigma(v)$} & \colhead{$\sigma_{\rm f}(\tau)$} & \colhead{$\sigma_{\rm f}(v)$}\\ \colhead{/ACT-CL} & \colhead{148 GHz\tablenotemark{a}} & \colhead{218 GHz\tablenotemark{a}} & \colhead{345 GHz} & \colhead{/arcmin} & \colhead{/keV} & \colhead{/keV}& \colhead{$/10^{-2}$} & \colhead{$/ \rm km\, s^{-1}$} & \colhead{$/10^{-3}$} & \colhead{$/ \rm km\, s^{-1}$} & \colhead{$/10^{-3}$} & \colhead{$/\rm km\, s^{-1}$}}
\startdata
J0102-4915 & 97 & 102& 183 & 2.50 & 14.5 & 1.0 & 3.1 & -1100 & 7.3 & 1800\tablenotemark{b} & 2.3 & 510\\
J0215-5212 & 40 & 42 & 152 & 3.16 & 5.9  & 1.3 & 2.8 & -1100 & 9.8 & 500\tablenotemark{b} & 5.9 & 350\\
J0232-5257 & 37 & 39 & 130 & 2.42 & 9.1  & 2.1 & 1.3 & -1200 & 6.3 & 1600\tablenotemark{b} & 3.0 & 810\\
J0330-5227 & 37 & 39 & 145 & 4.08 & 4.3  & 0.2\tablenotemark{b} & 1.9 & 100 & 8.4\tablenotemark{b} & 1000 & 2.0 & 280\\
J0438-5419 & 52 & 55 & 122 & 4.53 & 11.9 & 1.2 & 1.1 & 900   & 2.4 & 1000 & 1.1 & 530\\
J0546-5345 & 45 & 47 & 122 & 1.75 & 8.5  & 1.2\tablenotemark{b} & 5.2 & -300 & 13\tablenotemark{b} & 700 & 6.8 & 280\\
J0559-5249 & 42 & 44 & 213 & 3.08 & 8.1  & 0.8 & 1.3 & 3100  & 4.6 & 1500\tablenotemark{b} & 1.9 & 540\\
J0616-5227 & 42 & 44 & 160 & 2.60 & 6.6  & 0.8 & 3.6 & -300  & 8.3 & 600 & 4.5 & 260\\
J0658-5557 & 82 & 87 & 167 & 3.44 & 10.8 & 0.9 & 2.3 & 2500  & 5.1\tablenotemark{b} & 1000 & 2.1 & 450
\enddata
\tablenotetext{a}{Sensitivities estimated from the noise map of the ACT southern survey \citep{lindner_act_sens}, based on the position of each cluster}
\tablenotetext{b}{Where the reported upper and lower errors are different, we take the average, as we assume symmetrical Gaussian distributions}
\end{deluxetable*}

\section{Pixelization}
\label{sec:pixel}

\begin{figure}
\includegraphics[width=.5\columnwidth]{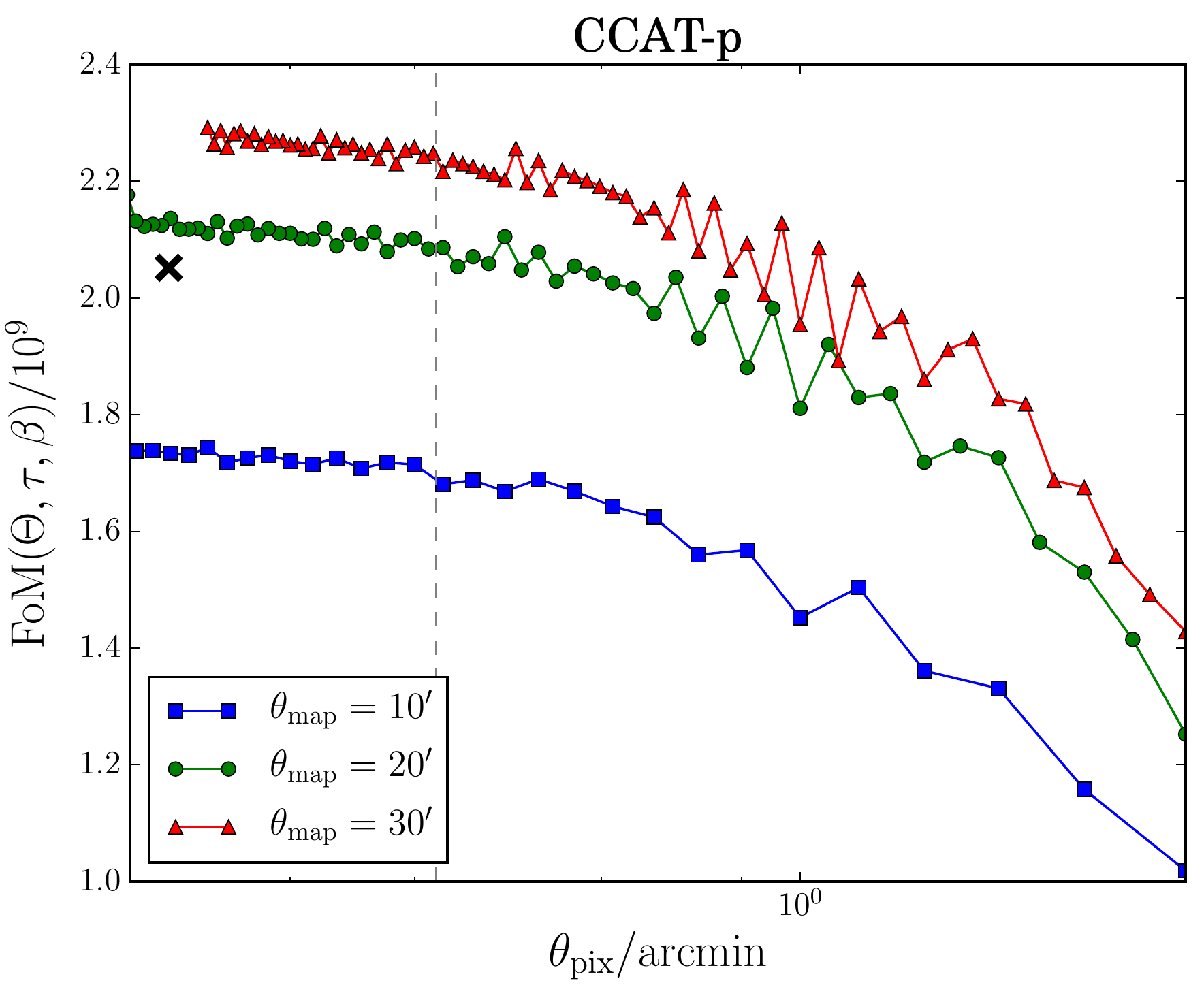}
\includegraphics[width=.5\columnwidth]{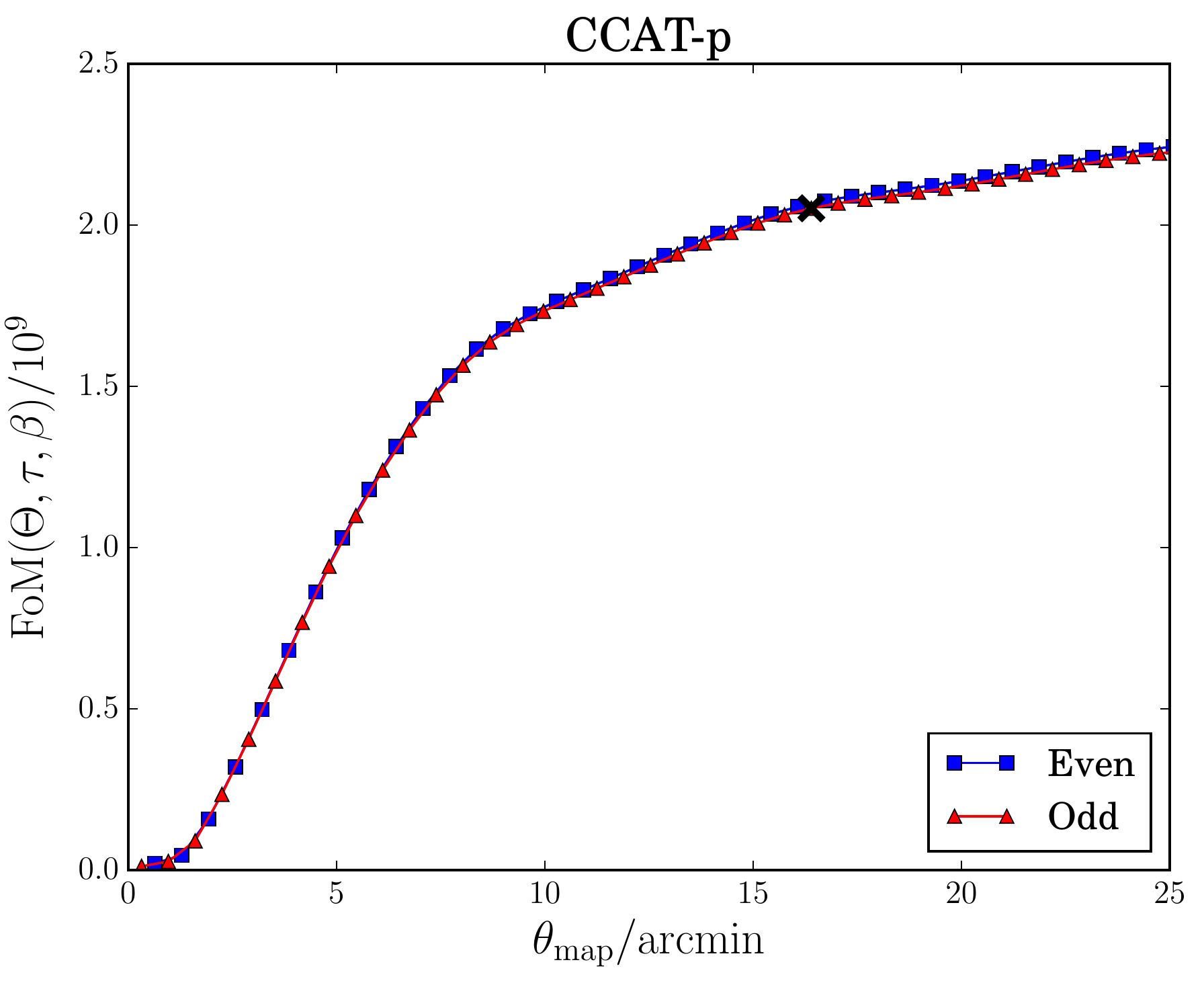} 
\caption{\textit{Left:} The variation of the FoM of all 3 parameters with the pixel size $\theta_\pix$ at three different map sizes for the CCAT$_{base}$ parameters. The plots are fully sampled (i.e. at every integer value of $s_\pix$). The vertical dashed line represents the FWHM of the smallest beam, and the black cross indicates the pixelization used throughout the paper, determined by equations \ref{pixsize_choice} and \ref{mapsize_choice}. \textit{Right:} The variation of the bounds with the map size $\theta_{\rm map}$ at the pixel size prescribed by equation \ref{pixsize_choice}. The black cross indicates the map size used throughout the paper, determined by equation \ref{mapsize_choice}. Since adding pixels can only improve the bounds, the odd and even pixel numbers must be monotonically increasing, which is why they are plotted separately. The cluster parameters used for these plots are described in Table~\ref{fiducial_cluster} ($\tfoo=3'$).
\label{pixsize_fig}}
\end{figure}

We have found that assumptions about pixelization and map size can have considerable impact on the FoM. We assume map pixels are on a square grid where each pixel is one observation as described above. Each pixel is a square of side $\theta_\pix = \sqrt{\Omega_\pix}$, and there are $s_\pix = \sqrt{n_\pix}$ pixels per side of the map, so that the map is a square of angular size $\theta_{\rm map} = s_\pix\theta_\pix$. The choice for $s_\pix$ and $\theta_\pix$ is not trivial and affects the signal-to-noise of the measurement.

Reducing the pixel size increases the instrumental noise per pixel and the galactic noise covariance between distinct pixels, due to $c^{\rm BS}$, while allowing a more accurate removal of the background terms, like the CMB. For a given beam size and instrumental noise there is hence an optimal trade-off between opposite effects. Explicitly, for a constant map size, decreasing the pixel size improves the signal-to-noise ratio up to a point, beyond which it plateaus. 

We study the dependence of the constraints on the pixel size choice for different map sizes. As an example, in Figure \ref{pixsize_fig} we plot the FoM results for CCAT$_{base}$ (for further discussion about these results see \S\ref{results}). As $\theta_\pix$ decreases, the FoM (\S \ref{fom}) plateaus towards a constant value. The point at which it approaches this constant value depends on the signal scale, $\tfoo$, and the beam widths $\xi$. Picking a small value for $\theta_\pix$ increases the computation time $t\propto \mathcal{O}(n_\pix^3) = \mathcal{O}(s_\pix^6)$. Thus we adopt the following rule of thumb to choose $\theta_\pix$ in the region where it plateaus, to minimise information loss while avoiding unnecessary computational complexity:
\begin{equation}\label{pixsize_choice}
\theta_\pix \equiv (\tfoo/10 + {\rm min}\, \xi ) / 2.5
\end{equation}

\begin{figure}
\includegraphics[width=0.5\columnwidth]{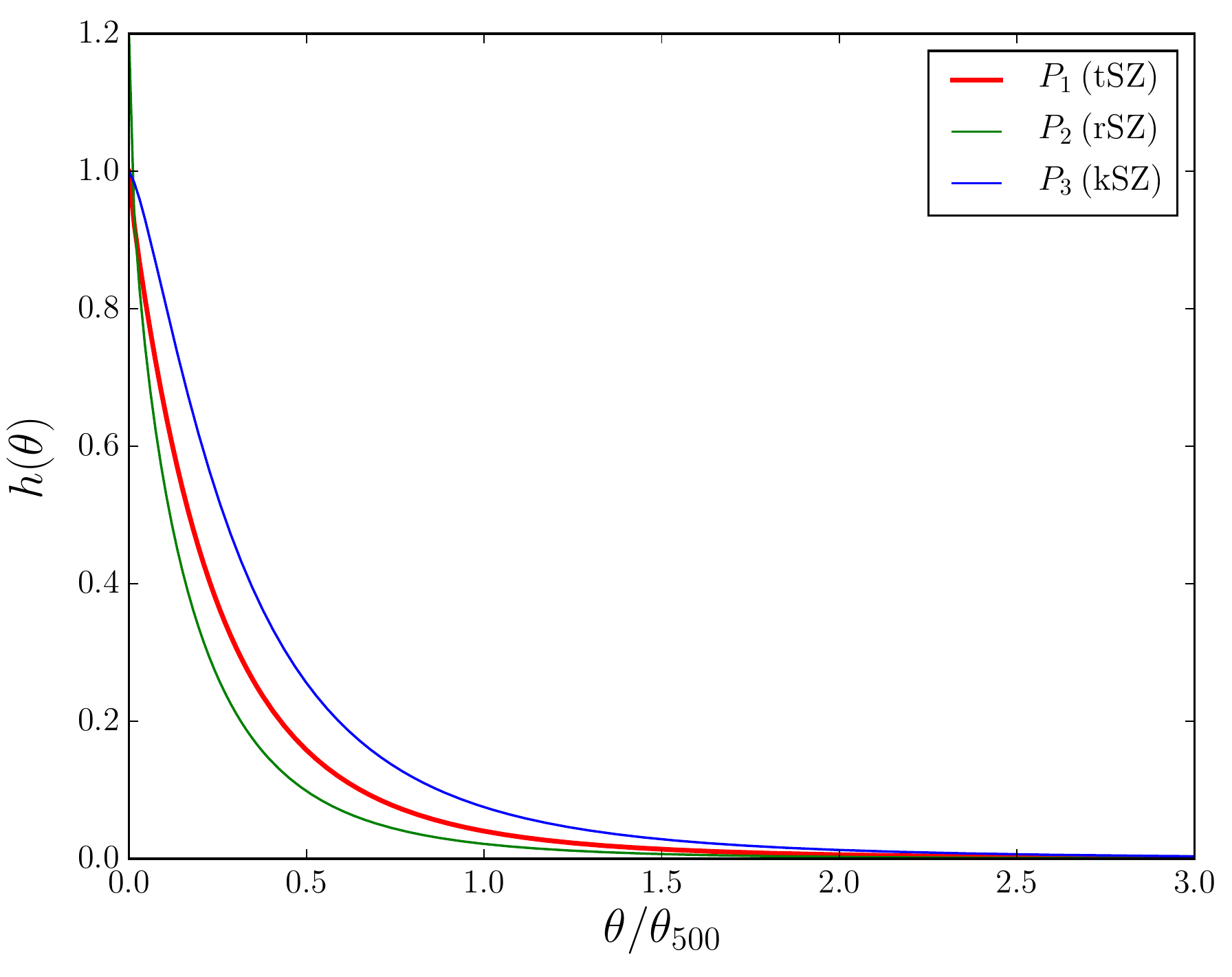}
\includegraphics[width=0.5\columnwidth]{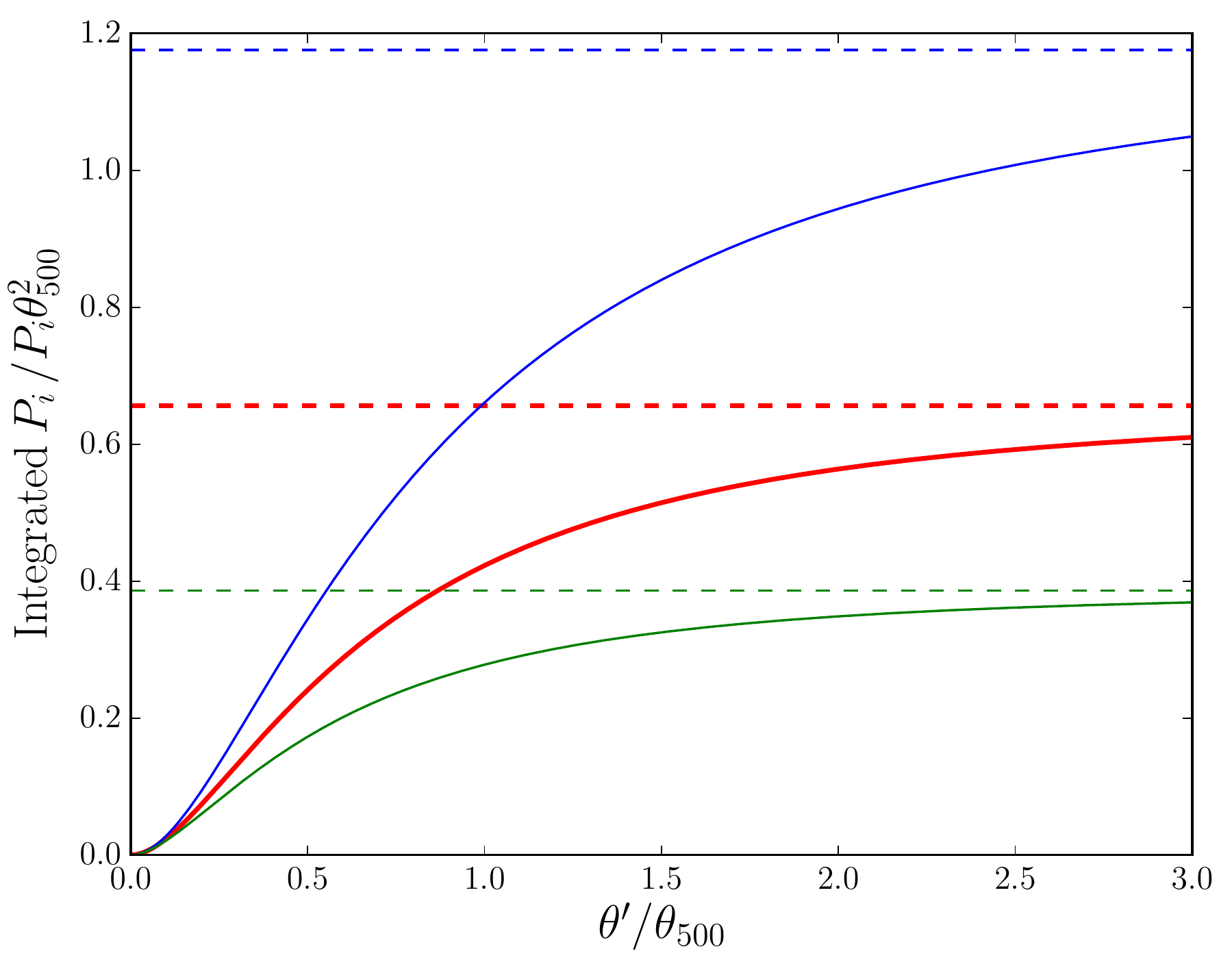}
\caption{
\textit{Left:} The cluster profiles of the various SZ signals described in \S\ref{SZ} as a function of angle from the cluster center $\theta$. The thick red line describes the tSZ profile $h(\theta)$ (ref. equation \ref{cluster_profile}), which is used in these forecasts for all three SZ components under the assumption of isothermality. The other lines describe the kSZ and relativistic correction to the tSZ components using the power-law temperature profile from \cite{2008A&A...486..359L}, to indicate how the signal profiles change when non-isothermality is considered. \textit{Right:}
The SZ parameters $P_i$ integrated within an aperture of radius $\theta'$, i.e. the total signal in a circular map of radius $\theta'$. The dashed horizontal lines represent the limit $\theta' \rightarrow \infty$, i.e. the total signal. The integrated tSZ parameter (thick red line) is the integrated comptonization parameter $Y(\theta')$ (equation \ref{int_y_eq}), which describes all three components under conditions of isothermality. The kSZ signal of a non-isothermal cluster \citep[like those in][]{2008A&A...486..359L} is greater than that of an isothermal cluster, while the rSZ signal is smaller.
\label{int_signal_plot}}
\end{figure}

The signal drops off sharply with distance (Figure~\ref{int_signal_plot}), so extending the map beyond the scale of $\tfoo$ yields little additional signal. The total signal within a radius $\theta'$ of the cluster center is also plotted in Figure~\ref{int_signal_plot}. Thus we expect the FoM to plateau out as we increase $\theta_{\rm map}$. The forecasted variation of the overall FoM with map size, with the pixel size held constant at the value prescribed by equation~\ref{pixsize_choice}, is plotted for CCAT$_{base}$ in Figure \ref{pixsize_fig}. We expect that the dependence of the results on the map size will become stronger if non-isothermal clusters are considered since the kSZ signal is more spread out (ref. Figure~\ref{int_signal_plot}).
In an attempt to standardize the results, we choose a constant radius $\theta = n\tfoo$ within which to capture the signal. A larger value for $n$ will minimize the variation due to the choice of pixelization, but also increase computational time and errors due to noise that has not been considered. Based on the signal dependence on $n$ (Figure 
\ref{int_signal_plot}), we set $n \equiv 2$. Accounting for beam convolution, this gives us the following rule of thumb for the map size:
\begin{equation}\label{mapsize_choice}
\rm \theta_{map} \equiv 2\,(2 \hspace{0.05 em} \tfoo + max\, \xi)
\end{equation}

The beam convolution with a map of a specific size presents another source of pixelization variation, as each frequency channel has a different beam size. For the map size selection, we use the maximum beam size to minimize loss of information.

\vspace{0.5in}

\bibliography{bib}

\end{document}